\documentstyle[prd,preprint,tighten,aps,eqsecnum,amssymb,amsbsy,newlfont,epsfig]{revtex}
\begin{document}
\draft
\preprint{
\begin{tabular}{r}
DFTT 18/96
\\
SISSA 35/96/EP
\\
UWThPh-1996-29
\\
hep-ph/9604364
\end{tabular}
}
\title{Short-baseline
neutrino oscillations
and
$\boldsymbol{(\beta\beta)}_{\boldsymbol{0\nu}}$-decay
in schemes
with an inverted mass spectrum}
\author{S.M. Bilenky}
\address{Joint Institute for Nuclear Research,
Dubna, Russia,
\\
and
\\
Institut f\"ur Theoretische Physik,
Universit\"at Wien,
Boltzmanngasse 5,
A-1090 Vienna, Austria.}
\author{C. Giunti}
\address{INFN, Sezione di Torino,
and Dipartimento di Fisica Teorica,
Universit\`a di Torino,
\\
Via P. Giuria 1, I-10125 Torino, Italy.}
\author{C.W. Kim}
\address{Department of Physics and Astronomy,
The Johns Hopkins University,
\\
Baltimore, Maryland 21218, USA,
\\
and
\\
Department of Physics,
Korea Advanced Institute of Science and Technology,
Daeduk, Korea.}
\author{S.T. Petcov}
\address{
Scuola Internazionale Superiore di Studi Avanzati,
I-34013 Trieste, Italy,
\\
and
\\
Institute of Nuclear Research and 
Nuclear Energy,
Bulgarian Academy of Sciences,
\\
BG-1784 Sofia, Bulgaria.}
\date{April 19, 1996}
\maketitle
\begin{abstract}
We have considered short-baseline neutrino oscillations,
$^3$H $\beta$-decay
and
$(\beta\beta)_{0\nu}$-decay
in two schemes with
an inverted mass spectrum
and mixing of three and four
massive neutrino fields.
We have analyzed the results
of the latest experiments on the search for
oscillations of terrestrial neutrinos
and we have discussed the compatibility
of the LSND indication in favor of neutrino oscillations
with the results of the other experiments.
In the framework of the models under consideration,
it is shown that
the observation of
$(\beta\beta)_{0\nu}$-decay
could allow to obtain information
about the CP violation in the lepton sector.
\end{abstract}

\pacs{12.15.Ff, 14.60.Lm, 14.60.Pq, 23.40.-s}

\narrowtext

\section{Introduction}
\label{intro}

The problem of neutrino mass and mixing
is the central issue
of today's neutrino physics.
The study of this problem
is associated with the search of
new scales in physics.
The investigation of
the neutrino mass problem
is also connected with the hopes to
understand the nature of
the dark matter in the universe.

At present there are several indications
in favor of non-zero neutrino mass
and mixing.
One indication comes from
the solar neutrino experiments.
In all four solar neutrino experiments
(Homestake \cite{Homestake},
Kamiokande \cite{Kamiokande},
GALLEX \cite{GALLEX}
and SAGE \cite{SAGE})
the observed event rates are significantly lower
than the event rates predicted
by the Standard Solar Models
(see Refs.\cite{bahcall,turk,cdf}).
After the test of the GALLEX and SAGE detectors
in special experiments with
radioactive $^{51}$Cr sources
\cite{GALLEXcal,SAGEcal}
the indications in favor of neutrino
oscillations coming from solar neutrino experiments have become
more significant.
Furthermore,
a phenomenological
analysis of the solar neutrino data
based on the assumption that
there is no neutrino mixing
indicates
\cite{phenomenological}
that the signals due 
to the $^7\mbox{Be}$ neutrinos
in the Cl-Ar
and Ga-Ge experiments are 
substantially lower than the signals
predicted by the Standard Solar Models.
No plausible astrophysical or nuclear physics 
explanation of this discrepancy has been proposed so far.

The solar neutrino data
can be explained by assuming neutrino mixing
with the MSW effect \cite{MSW,SOLMSW}
or vacuum oscillations
\cite{Pontecorvo,BP78,SOLVAC}.
In the simplest two-generation scheme,
with transitions between
$\nu_e$ and $\nu_{\mu}$
or
$\nu_e$ and $\nu_{\tau}$,
a fit of the data
with the MSW mechanism
yields the following values for the mixing parameters
\cite{KP95}:
$ \Delta m^2 \simeq
5 \times 10^{-6} \, \mbox{eV}^2 $
and
$ \sin^2 2\theta \simeq
7 \times 10^{-3} $
or
$ \Delta m^2 \simeq
2 \times 10^{-5} \, \mbox{eV}^2 $
and
$ \sin^2 2\theta
\simeq 0.8 $,
where
$ \Delta m^2 \equiv m_2^2 - m_1^2 $
($m_1$ and $m_2$ are the neutrino masses)
and $\theta$ is the mixing angle.
In the case of vacuum oscillations,
a fit of the data
yields
\cite{KP96}
$ \Delta m^2 \simeq
6 \times 10^{-11}  \, \mbox{eV}^2 $
and
$ \sin^2 2\theta \simeq 0.9 $.

Another indication in 
favor
of neutrino mixing
comes from the results of the experiments
on the detection of atmospheric neutrinos
\cite{Kamiokande-atmospheric,IMB,Soudan}.
In the Kamiokande, IMB and Soudan experiments
the ratio of $\mu$-like and $e$-like events
is less than expected.
The data can be explained by
$ \nu_\mu \leftrightarrows \nu_{\tau} $
or
$ \nu_\mu \leftrightarrows \nu_{e} $
oscillations
with
$ \Delta m^2 \simeq 10^{-2} \, \mbox{eV}^2 $
and a large mixing angle.

A third indication in favor of neutrino mixing
comes from the possible evidence of
$ \bar\nu_\mu \to \bar\nu_e $
transitions
reported recently by
the LSND collaboration
\cite{LSND}
(see, however, also Ref.\cite{hill}).
The explanation of the LSND data
in terms of
$ \bar\nu_\mu \leftrightarrows \bar\nu_e $
oscillations requires a value of
$ \Delta m^2 $
of the order of a few eV$^2$.

Finally,
neutrinos are very plausible candidates
for the dark matter in the universe
if their masses lie in the eV region
(see, for example, Refs.\cite{KT89,CWKim}).

All these indications in favor
of neutrino mass and mixing
could imply that
there are at least three different scales
for $ \Delta m^2 $.
In this case,
the number of massive neutrinos must be
more than three.
Let us notice that the results of the LEP experiments
which proved that the number of neutrino flavors is
equal to three
(see Ref.\cite{RPP})
do not constrain the
number of massive light neutrinos.
In the case of a Dirac neutrino mass term,
the total lepton number is conserved,
massive neutrinos are Dirac particles
and the number of massive neutrinos
can be equal to three.
In the case of a Dirac and Majorana
neutrino mass term,
which is typical for GUT models,
the total lepton number is not conserved,
massive neutrinos
are Majorana particles
and their number is larger than three.

We will consider
here short-baseline
oscillations of the terrestrial neutrinos 
in two possible schemes,
one with mixing
of three massive neutrino fields
and the other with mixing of four massive neutrino
fields.
In the case of three neutrinos
we will assume that the neutrino masses
satisfy the relation
$ m_1 \ll m_2 \simeq m_3 $.
In the case of four neutrinos
we will assume that
$ m_1 \simeq m_2 \ll m_3 \simeq m_4 $.
Models
with three and four massive neutrinos
with this types of mass relations
have been
considered in the recent articles
\cite{PS94,CM95,RS95}
and
\cite{CM93,PV93,SSF93,PHKC95,GCGG95,SG95}
respectively.
They are inspired by the experimental indications
in favor of neutrino mixing
and by astrophysical arguments
\cite{PHKC95}
in favor of the existence of two practically degenerate
neutrinos with masses in the eV range.
These schemes are compatible with
the constraints that follow from
the r-process production of heavy elements
in the neutrino-heated ejecta of supernovae
\cite{r-process}.

We will derive now the
general formulas for neutrino oscillations
in the framework of the schemes under consideration,
that will be used in the subsequent analyses.
Let us consider neutrino oscillations
in the case of two groups
of neutrinos with close masses.
We will assume that the masses of the
second group
are much larger than the masses
of the first group and
only one squared-mass difference
is relevant for oscillations of
the terrestrial neutrinos.

According to a general theory of neutrino mixing
(see, for example, Refs.\cite{BP78,BP87,CWKim}),
the left-handed flavor neutrino fields
$\nu_{{\alpha}L}$
are superpositions
of the left-handed components
of (Dirac or Majorana)
massive neutrino fields
$\nu_{i}$:
\begin{equation}
\nu_{{\alpha}L}
=
\sum_{i}
U_{{\alpha}i}
\nu_{iL}
\quad , \qquad
\alpha = e , \mu , \tau
\;.
\label{101}
\end{equation}
If the number of massive neutrinos is more than three,
we also have
\begin{equation}
\left( \nu_{sR} \right)^{c}
=
\sum_{i}
U_{si}
\nu_{iL}
\quad , \qquad
s = s_1 , s_2 , \ldots
\;,
\label{102}
\end{equation}
where 
$ \displaystyle
\left( \nu_{sR} \right)^{c}
=
\mathcal{C} \overline{\nu}_{sR}^{T}
$
($\mathcal{C}$ is the matrix of charge-conjugation),
$\nu_{sR}$
are right-handed (sterile) fields,
$U$ is a unitary mixing matrix
and the index $s$
denotes the last (one or more,
depending on the number of
the sterile fields) rows
of $U$.
From Eqs.(\ref{101}) and (\ref{102})
it follows that
not only transitions among active neutrino states
are possible,
but also transitions among active and sterile states.
We will consider the case of
two neutrino groups with
close masses,
\begin{equation}
m_1
\leq
\ldots
\leq
m_M
\qquad \mbox{and} \qquad
m_{M+1}
\leq
\ldots
\leq
m_N
\;,
\label{103}
\end{equation}
and we will assume that
\begin{equation}
m_j \ll m_k
\;,
\qquad \mbox{with} \qquad
\left\{
\begin{array}{l} \displaystyle
j=1,\ldots,M
\;,
\\ \displaystyle
k=M+1,\ldots,N
\;.
\end{array}
\right.
\label{104}
\end{equation}
We will also assume that
in the experiments with
terrestrial neutrinos we have
\arraycolsep=3pt
\begin{eqnarray}
{\displaystyle
\left( m^2_j - m^2_{j'} \right) L
\over\displaystyle
2 \, p
}
\ll 1
\;,
& \qquad \qquad &
j , j' \leq M
\;,
\label{105}
\\
{\displaystyle
\left( m^2_k - m^2_{k'} \right) L
\over\displaystyle
2 \, p
}
\ll 1
\;,
& \qquad \qquad &
k , k' > M
\;,
\label{106}
\end{eqnarray}
where $L$ is the distance
between the source and the detector
and $p$ is the neutrino momentum.
Taking into account
Eqs.(\ref{105}) and (\ref{106}),
the amplitude of
$ \nu_{\alpha} \to \nu_{\beta} $
transitions
(here $\nu_{\alpha}$ and $\nu_{\beta}$
are
the active or sterile neutrinos)
for the
short-baseline terrestrial experiments
is given by
\arraycolsep=0pt
\begin{eqnarray}
\mathcal{A}_{\nu_{\alpha}\to\nu_{\beta}}
&=&
\mathrm{e}^{ - i E_1 t }
\sum_{j=1}^{N}
U_{\beta j}
\,
U_{\alpha j}^{*}
\,
\mathrm{e}^{ - i \left( E_j - E_1 \right) t }
\nonumber
\\
&\simeq&
\mathrm{e}^{ - i E_1 t }
\left\{
\sum_{j=1}^{M}
U_{{\beta}j}
U_{{\alpha}j}^{*}
+
\exp
\left(
- i
{\displaystyle
\Delta m^2 L
\over\displaystyle
2 p
}
\right)
\sum_{k=M+1}^{N}
U_{{\beta}k}
U_{{\alpha}k}^{*}
\right\}
\;,
\label{107}
\end{eqnarray}
with
$ \Delta m^2 \equiv m^2_N - m^2_1 $,
$ \displaystyle
E_i
=
\sqrt{ p^2 + m_i^2 }
\simeq
p
+
m_i^2
/
2 p
$
and
$ t \simeq L $.
Furthermore,
using the unitarity relation
\begin{equation}
\sum_{j=1}^{M}
U_{{\beta}j}
U_{{\alpha}j}^{*}
=
\delta_{\alpha\beta}
-
\sum_{k=M+1}^{N}
U_{{\beta}k}
U_{{\alpha}k}^{*}
\;,
\label{108}
\end{equation}
we can rewrite the transition amplitude
$\mathcal{A}_{\nu_{\alpha}\to\nu_{\beta}}$
in the form
\begin{equation}
\mathcal{A}_{\nu_{\alpha}\to\nu_{\beta}}
\simeq
\mathrm{e}^{ - i E_1 T }
\left\{
\delta_{\alpha\beta}
+
\left[
\exp
\left(
- i
{\displaystyle
\Delta m^2 L
\over\displaystyle
2 p
}
\right)
-
1
\right]
\sum_{k=M+1}^{N}
U_{{\beta}k}
U_{{\alpha}k}^{*}
\right\}
\;.
\label{109}
\end{equation}
From Eq.(\ref{109}),
for the probability of
$ \nu_{\alpha} \to \nu_{\beta} $
transitions
with $\beta\not=\alpha$
we obtain the following expression:
\begin{equation}
P_{\nu_{\alpha}\to\nu_{\beta}}
=
{1\over2}
\,
A_{\nu_{\alpha};\nu_{\beta}}
\left(
1
-
\cos
{\displaystyle
\Delta m^2 \, L
\over\displaystyle
2 \, p
}
\right)
\;,
\label{110}
\end{equation}
where
\begin{equation}
A_{\nu_{\alpha};\nu_{\beta}}
=
4
\left|
\sum_{k=M+1}^{N}
U_{{\beta}k}
U_{{\alpha}k}^{*}
\right|^2
\label{111}
\end{equation}
is the
amplitude of
$ \nu_{\alpha} \to \nu_{\beta} $
oscillations.

The expression for the
survival probability of
$\nu_{\alpha}$
can be obtained from Eq.(\ref{110})
and the conservation of the total probability:
\begin{equation}
P_{\nu_{\alpha}\to\nu_{\alpha}}
=
1
-
\sum_{\beta\not=\alpha}
P_{\nu_{\alpha}\to\nu_{\beta}}
=
1
-
{1\over2}
\,
B_{\nu_{\alpha};\nu_{\alpha}}
\left(
1
-
\cos
{\displaystyle
\Delta m^2 \, L
\over\displaystyle
2 \, p
}
\right)
\;,
\label{112}
\end{equation}
where
the oscillation amplitude
$ B_{\nu_{\alpha};\nu_{\alpha}} $
is given by
\begin{equation}
B_{\nu_{\alpha};\nu_{\alpha}}
=
\sum_{\beta\not=\alpha}
A_{\nu_{\alpha};\nu_{\beta}}
\;.
\label{113}
\end{equation}
Using the unitarity of the mixing matrix,
from Eqs.(\ref{111}) and (\ref{113})
we obtain
\begin{equation}
B_{\nu_{\alpha};\nu_{\alpha}}
=
4
\left(
\sum_{k=M+1}^{N}
\left| U_{{\alpha}k} \right|^2
\right)
\left(
1
-
\sum_{k=M+1}^{N}
\left| U_{{\alpha}k} \right|^2
\right)
\;.
\label{114}
\end{equation}

Let us notice that
from Eq.(\ref{111}) we have
\begin{equation}
A_{\nu_{\alpha};\nu_{\beta}}
=
A_{\nu_{\beta};\nu_{\alpha}}
\;.
\label{115}
\end{equation}
From Eqs.(\ref{110}) and (\ref{115})
it follows that
\begin{equation}
P_{\nu_{\alpha}\to\nu_{\beta}}
=
P_{\nu_{\beta}\to\nu_{\alpha}}
\;.
\label{116}
\end{equation}
Due to CPT invariance we have
(see, for example, Refs.\cite{BP78,BP87,CWKim})
\begin{equation}
P_{\nu_{\alpha}\to\nu_{\beta}}
=
P_{\bar\nu_{\beta}\to\bar\nu_{\alpha}}
\;.
\label{117}
\end{equation}
Thus,
if only one squared-mass difference
between the heaviest and lightest neutrinos
is relevant for the oscillations
of the terrestrial neutrinos,
even in the case of non-conservation
of CP in the lepton sector,
the transition probabilities of neutrinos
($\nu_{\alpha}\to\nu_{\beta}$)
and antineutrinos
($\bar\nu_{\alpha}\to\bar\nu_{\beta}$)
are equal:
\begin{equation}
P_{\nu_{\alpha}\to\nu_{\beta}}
=
P_{\bar\nu_{\alpha}\to\bar\nu_{\beta}}
\;.
\label{118}
\end{equation}

Before closing this Section,
let us remark that
from the unitarity relation (\ref{108})
it follows that
the amplitude (\ref{111})
of
$\nu_{\alpha}\leftrightarrows\nu_{\beta}$
oscillations
can also be written as
\begin{equation}
A_{\nu_{\alpha};\nu_{\beta}}
=
4
\left|
\sum_{j=1}^{M}
U_{{\beta}j}
U_{{\alpha}j}^{*}
\right|^2
\;.
\label{119}
\end{equation}
In an analogous way,
from Eqs.(\ref{108})
it follows that
the oscillation amplitudes
$ B_{\nu_{\alpha};\nu_{\alpha}} $
given in Eq.(\ref{114})
can also be written as
\begin{equation}
B_{\nu_{\alpha};\nu_{\alpha}}
=
4
\left(
\sum_{j=1}^{M}
\left| U_{{\alpha}j} \right|^2
\right)
\left(
1
-
\sum_{j=1}^{M}
\left| U_{{\alpha}j} \right|^2
\right)
\;.
\label{120}
\end{equation}
Thus,
the expressions for the
oscillation amplitudes
$ A_{\nu_{\alpha};\nu_{\beta}} $
and
$ B_{\nu_{\alpha};\nu_{\alpha}} $
can be written with the sum over
the indexes of the neutrinos of the second group
(Eqs.(\ref{111}) and (\ref{114}))
or with the sum over
the indexes of the neutrinos of the first group
(Eqs.(\ref{119}) and (\ref{120})).

\section{Mixing of three massive neutrinos}
\label{SN3}

Let us consider first
the schemes with mixing of three massive neutrino
fields.
The case of a neutrino mass hierarchy
\begin{equation}
m_1 \ll m_2 \ll m_3
\;.
\label{201}
\end{equation}
was considered recently
in Refs.\cite{PS94,Lisi,BBGK95,BBGK96,Minakata,BPW95,CF96}.
In this case
the oscillations of the terrestrial neutrinos
are determined by three parameters:
$ \Delta m^2 $
and the two mixing parameters
$|U_{e3}|^2$
and
$|U_{\mu3}|^2$.
In Ref.\cite{BBGK95,BBGK96}
it was shown that,
if a hierarchy of couplings is realized in the lepton sector
(i.e. $|U_{e3}|^2$
and
$|U_{\mu3}|^2$
are small),
the LSND result is not compatible
with the results of all the other
reactor and accelerator experiments
on the search for neutrino oscillations.
Instead,
the result of the LSND experiment is
compatible with the results of all
the other experiments
in the scheme with
the neutrino mass hierarchy (\ref{201})
if
$|U_{e3}|^2$ is small
and
$|U_{\mu3}|^2$ is large (close to one).
If massive neutrinos are Majorana particles,
neutrinoless double-beta decay
is allowed and for
$ \Delta m^2 \gtrsim 5 \, \mbox{eV}^2 $
this process can have a rate
in the region of sensitivity
of the next generation of experiments
\cite{PS94,BBGK96}.  

Here we will consider
another possible scheme with three
massive neutrinos
and one
$ \Delta m^2 $
relevant for the oscillations of
the terrestrial neutrinos.
We will assume the following
relation among the three neutrino masses:
\begin{equation}
m_1 \ll m_2 \simeq m_3
\label{202}
\end{equation}
This scheme was recently discussed
in Refs.\cite{PS94,CM95,RS95}.
In favor of such a scheme
there are some cosmological arguments
\cite{PHKC95}
and some astrophysical arguments
concerning the r-process production of
heavy elements in the
neutrino-heated ejecta of supernovae
\cite{r-process}.

We will assume that the squared-mass difference
$ \Delta m^2_{32} \equiv m^2_3 - m^2_2 $
is small and is relevant for
the suppression of the flux of solar $\nu_e$'s on the earth.
In this case,
from the general formulas
given in Section \ref{intro}
it follows that
the oscillations of the terrestrial neutrinos
are characterized by the values of
$ \Delta m^2 \equiv m^2_3 - m^2_1 $
and two mixing parameters
$ |U_{e1}|^2 $ and $ |U_{\mu1}|^2 $.
The probabilities of
$ \nu_{\alpha} \to \nu_{\beta} $
($\beta\not=\alpha$)
and
$ \nu_{\alpha} \to \nu_{\alpha} $
transitions
are given by Eqs.(\ref{110}) and (\ref{112})
with the following oscillation amplitudes:
\arraycolsep=0pt
\begin{eqnarray}
&&
A_{\nu_{\alpha};\nu_{\beta}}
=
4
|U_{\beta1}|^2
|U_{\alpha1}|^2
\;,
\label{203}
\\
&&
B_{\nu_{\alpha};\nu_{\alpha}}
=
4
|U_{\alpha1}|^2
\left(
1
-
|U_{\alpha1}|^2
\right)
\;.
\label{204}
\end{eqnarray}

Let us consider first the constraints
in the scheme under consideration
following from the results of
reactor and accelerator disappearance experiments.
We will use the exclusion plots
obtained
in the
$ \bar\nu_e \to \bar\nu_e $
Bugey reactor experiment
\cite{Bugey95}
and in the
$ \nu_\mu \to \nu_\mu $
CDHS and CCFR84 accelerator experiments
\cite{CDHS84,CCFR84}.
At fixed values of
$ \Delta m^2 $,
the allowed values of
the oscillation amplitudes
$ B_{\nu_{e};\nu_{e}} $
and
$ B_{\nu_{\mu};\nu_{\mu}} $
are constrained by
\begin{equation}
B_{\nu_{\alpha};\nu_{\alpha}}
\le
B_{\nu_{\alpha};\nu_{\alpha}}^{0}
\qquad
(\alpha=e,\mu)
\;.
\label{205}
\end{equation}
The values of
$ B_{\nu_{e};\nu_{e}}^{0} $
and
$ B_{\nu_{\mu};\nu_{\mu}}^{0} $
can be obtained from the corresponding exclusion curves.
We will consider values of
$ \Delta m^2 $
in the interval
\begin{equation}
0.5 \, \mathrm{eV}^2 \lesssim
\Delta m^2
\lesssim 10^{2} \, \mathrm{eV}^2
\;,
\label{288}
\end{equation}
that covers the range
where positive indications
in favor of
$ \bar\nu_\mu \leftrightarrows \bar\nu_e $
oscillations
were reported by the LSND collaboration
\cite{LSND}.
In this range of $ \Delta m^2 $
the quantities
$ B_{\nu_{e};\nu_{e}}^{0} $
and
$ B_{\nu_{\mu};\nu_{\mu}}^{0} $
are smaller than 0.16 and 0.40,
respectively.
From Eq.(\ref{204}) it follows that
the parameters
$ \left| U_{\alpha1} \right|^2 $
at fixed values of
$ \Delta m^2 $
must satisfy one of the following inequalities:
\arraycolsep=3pt
\begin{eqnarray}
&&
\left| U_{\alpha1} \right|^2
\le
a^{0}_{\alpha}
\label{206}
\\
\mbox{or}
&&
\nonumber
\\
&&
\left| U_{\alpha1} \right|^2
\ge
1 - a^{0}_{\alpha}
\;,
\label{207}
\end{eqnarray}
with
\begin{equation}
a^{0}_{\alpha}
\equiv
{1\over2}
\left(
1
-
\sqrt{ 1 - B_{\nu_{\alpha};\nu_{\alpha}}^{0} }
\right)
\;.
\label{208}
\end{equation}
In the interval of
$ \Delta m^2 $
given in Eq.(\ref{288})
the values of
$a^{0}_{e}$
and
$a^{0}_{\mu}$
are smaller than 0.04 and 0.1,
respectively.
Thus,
in the wide range of
$ \Delta m^2 $
under consideration
the mixing parameters
$ \left| U_{e1} \right|^2 $
and
$ \left| U_{\mu1} \right|^2 $
can either be small
or large
(close to one).

We have assumed that
$ \Delta m^2_{32} $
is relevant for the suppression of the flux
of solar neutrinos on the earth
and that
$ \Delta m^2_{32} \ll \Delta m^2 $.
In this case
the survival probability of the solar $\nu_e$'s
is given by
\begin{equation}
P_{\nu_e\to\nu_e}
=
\left(
1
-
\left| U_{e1} \right|^2
\right)^2
P_{\nu_e\to\nu_e}^{(2,3)}
+
\left| U_{e1} \right|^4
\;,
\label{209}
\end{equation}
where
$ P_{\nu_e\to\nu_e}^{(2,3)} $
is the survival probability
due to the mixing of $\nu_e$
with $\nu_2$ and $\nu_3$
(Eq.(\ref{209}) can be obtained with
the method presented in Ref.\cite{SS92}).
From Eq.(\ref{209})
and from the results of
the Bugey disappearance experiment
it follows that,
in the case of a large
$ \left| U_{e1} \right|^2 $,
we have
$
P_{\nu_e\to\nu_e}
\ge
0.92
$
for all values of the neutrino energy.
Such a large lower bound of the $\nu_e$ survival probability
is not compatible
with the results of solar neutrino experiments
\cite{Homestake,Kamiokande,GALLEX,SAGE,phenomenological}.

It is obvious that,
due to the unitarity constraint
$ \left| U_{e1} \right|^2 +
\left| U_{\mu1} \right|^2 \leq 1 $,
the mixing parameters
$ \left| U_{e1} \right|^2 $
and
$ \left| U_{\mu1} \right|^2 $
cannot be both large.
Thus,
from the results of reactor and accelerator
disappearance experiments
and
from the results of solar neutrino experiments
it follows that the values of the parameters
$ \left| U_{e1} \right|^2 $
and
$ \left| U_{\mu1} \right|^2 $
can lie in one of the following two regions:

\def\theenumi{\Roman{enumi}}

\begin{enumerate}

\item \label{N3R1}
The region of small
$ \left| U_{e1} \right|^2 $
and
$ \left| U_{\mu1} \right|^2 $.

\item \label{N3R2}
The region of small
$ \left| U_{e1} \right|^2 $
and large
$ \left| U_{\mu1} \right|^2 $.

\end{enumerate}

The fact that the parameter
$ \left| U_{e1} \right|^2 $
is small
can have important implications
for $^3$H $\beta$-decay experiments
and
for the experiments on the search
for neutrinoless double-beta decay
($(\beta\beta)_{0\nu}$-decay).

The electron spectrum in the decay
$ \mbox{}^3\mbox{H} \to
\mbox{}^3\mbox{He} + e^{-} + \bar\nu_e $
is given by
\begin{equation}
{\displaystyle
\mathrm{d} N
\over\displaystyle
\mathrm{d} E
}
=
C \, p_e \, E_e
\left( Q - T \right)
F(E_e)
\sum_{i}
\left| U_{ei} \right|^2
\sqrt{
\left( Q - T \right)^2
-
m^2_i
}
\;.
\label{210}
\end{equation}
Here Q is the energy release,
$p_e$ and $E_e$
are the electron momentum and energy,
$ T = E_e - m_e $,
$F(E_e)$ is the Fermi function
and $C$ is a constant.
Taking into account that
$ \left| U_{e1} \right|^2 \ll 1 $
and
$ m_2 \simeq m_3 $,
from Eq.(\ref{210}) we have
\begin{equation}
{\displaystyle
\mathrm{d} N
\over\displaystyle
\mathrm{d} E
}
=
C \, p_e \, E_e
\left( Q - T \right)
F(E_e) \,
\sqrt{
\left( Q - T \right)^2
-
m^2_3
}
\;.
\label{211}
\end{equation}
This is the usual expression
for the $^3$H $\beta$-decay spectrum
from which information
about the value of the ``electron neutrino mass''
$ m_{\nu_{e}} $
is extracted.
The upper limit for $ m_{\nu_{e}} $ obtained
in the latest experiment
\cite{troitsk}
is
$ m_{\nu_{e}} \leq 4.5 \, \mbox{eV} $,
which implies that
\begin{equation}
\Delta m^2 \lesssim 20 \, \mbox{eV}^2
\label{212}
\end{equation}

If $\nu_2$ and $\nu_3$ contribute
to the dark matter in the universe,
the masses $ m_2\simeq m_3 $
are in the eV range.
Thus,
further improvements of the sensitivity of
$^3$H $\beta$-decay experiments
could have important implications
for the dark matter problem.

If massive neutrinos are Majorana particles,
neutrinoless double-beta decay
of some even-even nuclei is allowed.
The matrix element of
$(\beta\beta)_{0\nu}$ decay
is proportional to
\cite{mbb}
\begin{equation}
\langle m \rangle
=
\sum_{i}
U_{ei}^2
\,
m_{i}
\;.
\label{213}
\end{equation}
From the results of the experiments
on the search for
$(\beta\beta)_{0\nu}$ decay
it follows that
$ \left| \langle m \rangle \right| \lesssim 1-2 \, \mbox{eV} $
(see the review in Ref.\cite{MV94}).
The expected sensitivity
of experiments of the next generation is
$ \left| \langle m \rangle \right| \simeq $
a few
$ 10^{-1} \, \mbox{eV} $
(see Ref.\cite{Moe95}).

In the model under consideration
\begin{equation}
\langle m \rangle
\simeq
\left( U_{e2}^2 + U_{e3}^2 \right)
m_{3}
\;.
\label{214}
\end{equation}
We will consider the
general case of non-conservation of CP
in the lepton sector.
The case of CP conservation
was already discussed in Ref.\cite{PS94}.

Let us write $U_{ek}$ as
\begin{equation}
U_{ek}
=
|U_{ek}| \, \mathrm{e}^{i\phi_{k}}
\;.
\label{215}
\end{equation}
If CP is conserved in the lepton sector we have
\cite{CPparity}
\begin{equation}
U_{ek}
=
U_{ek}^{*} \, \eta_{k}
\;,
\label{216}
\end{equation}
where $ \eta_{k} = \pm i $ is the CP parity
of the Majorana neutrino with mass $m_k$.
From Eqs.(\ref{215}) and (\ref{216})
it follows that
in the case of CP invariance we have
\begin{equation}
\phi_{k} = \pm { \pi \over 4 }
\;.
\label{217}
\end{equation}
Furthermore,
taking into account that
$ |U_{e1}|^2 \ll 1 $,
from the unitarity of the mixing matrix we have
\begin{equation}
|U_{e2}|^2 + |U_{e3}|^2 \simeq 1
\;.
\label{218}
\end{equation}
From Eqs.(\ref{214}), (\ref{215}) and (\ref{218})
we obtain
\begin{equation}
\left| \langle m \rangle \right|
\simeq
m_3
\sqrt{
1
-
4
\,
|U_{e2}|^2
\,
|U_{e3}|^2
\,
\sin^2 \Phi
}
\;,
\label{219}
\end{equation}
with
\begin{equation}
\Phi \equiv \left| \phi_3 - \phi_2 \right|
\;.
\label{220}
\end{equation}
If CP is conserved,
we have
$ \Phi = 0 $ or $\pi/2$
for the same or opposite
CP parities of $\nu_2$ and $\nu_3$,
respectively.

Due to Eq.(\ref{218}) we can write
\begin{equation}
|U_{e2}| \simeq \cos\theta
\qquad \mbox{and} \qquad
|U_{e3}| \simeq \sin\theta
\;.
\label{221}
\end{equation}
With the help of Eqs.(\ref{219}) and (\ref{221})
we obtain
\begin{equation}
{\displaystyle
\left| \langle m \rangle \right|^2
\over\displaystyle
m_3^2
}
\simeq
1
-
\sin^2 2\theta
\,
\sin^2 \Phi
\;.
\label{222}
\end{equation}
From Eq.(\ref{222}) it follows that
\begin{equation}
1
-
\sin^2 2\theta
\leq
{\displaystyle
\left| \langle m \rangle \right|^2
\over\displaystyle
m_3^2
}
\leq
1
\;.
\label{223}
\end{equation}
The boundary values of
$ \left| \langle m \rangle \right|^2 / m_3^2 $
correspond to CP conservation in the lepton sector:
the upper (lower) bound corresponds
to the case of equal (opposite) CP parities
of $\nu_2$ and $\nu_3$.

Information about the value of the parameter
$ \sin^2 2\theta $
can be obtained from the results of
solar neutrino experiments.
If
$ \sin^2 2\theta \ll 1 $,
which corresponds to the MSW solution
with a small mixing angle,
we have
\begin{equation}
\left| \langle m \rangle \right| \simeq m_3
\;,
\label{224}
\end{equation}
independently from the value of
$\Phi$
and from the conservation of CP.
In the case of a large value of the parameter
$ \sin^2 2\theta $,
which correspond to the MSW solution
with a large mixing angle
or to the vacuum oscillation solution,
in the future
it will be possible to obtain information
about
the violation of
CP in the lepton sector
if both the
$^3$H $\beta$-decay
and
$(\beta\beta)_{0\nu}$ decay
experiments
will obtain positive results.
In fact,
from the measurement of
$ m_3 $ in $^3$H $\beta$-decay experiments,
of
$ \left| \langle m \rangle \right| $
in $(\beta\beta)_{0\nu}$ decay
experiments,
and of
$ \sin^2 2\theta $
in the solar neutrino experiments,
with the help of Eq.(\ref{222})
it will be possible to determine the value of
$ \sin^2 \Phi $
(if $ \sin^2 2\theta $ is large).
Let us emphasize that
in the case of CP conservation
the relative CP parities of $\nu_2$ and $\nu_3$
can be determined:
the CP parities are equal
if $ \sin^2 \Phi = 0 $
and opposite if
$ \sin^2 \Phi = 1 $.

Now we will discuss the implications
in the model under consideration
of the results of the appearance neutrino oscillation
experiments.
We will use the exclusion plots,
that were obtained in the BNL E776 experiment
\cite{BNLE776}
searching for
$ \nu_\mu \to \nu_e $
transitions,
in the FNAL E531
and
CCFR95
experiments
\cite{FNALE531,CCFR95}
searching for
$ \nu_\mu \to \nu_\tau $
transitions
and
in the FNAL E531 experiment
\cite{FNALE531}
searching for
$ \nu_e \to \nu_\tau $
transitions.
We can use the results
obtained in Ref.\cite{BBGK95,BBGK96}
for the case of mixing of three neutrinos
and the mass hierarchy (\ref{201})
if we make the change
\begin{equation}
|U_{\alpha3}|^2 \to |U_{\alpha1}|^2
\;.
\label{225}
\end{equation}
In the following two subsections
we will briefly present the main results
for the two allowed regions
of the values of the parameters
$|U_{e1}|^2$ and $|U_{\mu1}|^2$.

\subsection{The region \protect\ref{N3R1}
of small
$\boldsymbol{|U_{\boldsymbol{e1}}|^2}$
and
$\boldsymbol{|U_{\boldsymbol{\mu1}}|^2}$}
\label{SN3R1}

From Eq.(\ref{203}) it follows that
the oscillation amplitude
$ A_{\nu_{\mu};\nu_{e}} $
is quadratic in the small quantities
$|U_{e1}|^2$ and $|U_{\mu1}|^2$,
whereas the oscillation amplitudes
$ A_{\nu_{\mu};\nu_{\tau}} $
and
$ A_{\nu_{e};\nu_{\tau}} $
depend only linearly on these quantities.
Thus,
we expect that in the region \ref{N3R1}
$ \nu_\mu \leftrightarrows \nu_e $
oscillations are suppressed.

From Eq.(\ref{203}),
at fixed values of
$ \Delta m^2 $
in the interval under consideration,
we have
\begin{equation}
A_{\nu_{\mu};\nu_{e}}
\le
4 \, a^{0}_{e} \, a^{0}_{\mu}
\;,
\label{226}
\end{equation}
with
$a^{0}_{e}$ and $a^{0}_{\mu}$
given by Eq.(\ref{208}).
In Fig.\ref{n3r1muel}
we have plotted the curve
that represents this
upper bound obtained
from the results
of the Bugey,
CDHS and CCFR84 experiments
(the curve passing through the circles).
In Fig.\ref{n3r1muel}
we have also plotted the exclusion curves
obtained in the
BNL E776
\cite{BNLE776}
(dash-dotted line)
and
KARMEN
\cite{KARMEN}
(dash-dot-dotted line)
experiments
on the search of
$ \nu_\mu \to \nu_e $
transitions.
The region allowed by the results of the LSND experiment
is shown in Fig.\ref{n3r1muel}
as the shadowed region between the two solid lines.
Taking into account that
$
A_{\nu_{\mu};\nu_{e}}
\le
B_{\nu_{e};\nu_{e}}
$
(see Eq.(\ref{113})),
we also plotted in Fig.\ref{n3r1muel}
the exclusion curve
for
$ B_{\nu_{e};\nu_{e}} $
found in the Bugey experiment
(dashed line).
It can be seen from the figure that,
for small values of
$ \Delta m^2 $
($ \Delta m^2 \lesssim 0.5 \, \mbox{eV}^2 $),
this bound on
$ A_{\nu_{\mu};\nu_{e}} $
is stronger than the direct bound
obtained by the BNL E776 and KARMEN experiments.
It is also clear from the figure that
this bound is not compatible with the result
of the LSND experiment
for
$ \Delta m^2 \lesssim 0.2 \, \mbox{eV}^2 $.

It can be seen from Fig.\ref{n3r1muel}
that,
in the range of
$ \Delta m^2 $
under consideration,
with the exception of the region
$ 10 \, \mbox{eV}^2
\lesssim \Delta m^2 \lesssim
60 \, \mbox{eV}^2 $,
the limits on the oscillation amplitude
$ A_{\nu_{\mu};\nu_{e}} $
that can be obtained
from the results of disappearance experiments
are more stringent than the limits
obtained in the direct experiments searching for
$ \nu_\mu \to \nu_e $
transitions.

For
$ \Delta m^2 \gtrsim 4 \, \mbox{eV}^2 $
we can obtain even stronger limits
on the oscillation amplitude
$ A_{\nu_{\mu};\nu_{e}} $
if we take into account
the results of the FNAL E531 and CCFR95 experiments
on the search for
$ \nu_\mu \to \nu_\tau $
transitions.
In fact,
in the linear approximation over the small quantities
$|U_{e1}|^2$ and $|U_{\mu1}|^2$,
we have
\begin{equation}
\left| U_{\mu1} \right|^2
\simeq
{\displaystyle
A_{\nu_{\mu};\nu_{\tau}}
\over\displaystyle
4
}
\;.
\label{227}
\end{equation}
From Eqs.(\ref{203}) and (\ref{227})
we obtain the following upper bound:
\begin{equation}
A_{\nu_{\mu};\nu_{e}}
\lesssim
a^{0}_{e}
\,
A_{\nu_{\mu};\nu_{\tau}}^{0}
\;,
\label{228}
\end{equation}
where
$ A_{\nu_{\mu};\nu_{\tau}}^{0} $
is the upper bound
for the amplitude of
$ \nu_{\mu} \leftrightarrows \nu_{\tau} $
oscillations given by the exclusion curves
obtained in the FNAL E531 and CCFR95 experiments.
The bound (\ref{228}) on the oscillation amplitude
$ A_{\nu_{\mu};\nu_{e}} $
obtained
from the results of the Bugey, FNAL E531 and CCFR95 experiments
is presented in Fig.\ref{n3r1muel}
(the curve passing through the triangles).

From Fig.\ref{n3r1muel}
it can be seen that
the results of reactor and accelerator
disappearance experiments,
together with the results of
$ \nu_\mu \to \nu_\tau $
appearance experiments
allow us to exclude practically
all the region of the parameters
$ \Delta m^2 $
and
$ A_{\nu_{\mu};\nu_{e}} $
that is allowed by the LSND experiment.
Let us stress that this result is valid
if the parameters
$ \left| U_{e1} \right|^2 $
and
$ \left| U_{\mu1} \right|^2 $
are both small.

In Fig.\ref{n3r1muel},
we have also shown the region in the
$ A_{\nu_{\mu};\nu_{e}} $--$ \Delta m^2 $
plane that will be explored
when the projected sensitivity of the
CHORUS \cite{CHORUS} and NOMAD \cite{NOMAD}
experiments,
which are searching for
$ \nu_\mu \to \nu_\tau $
transitions,
is reached
(the region limited by the line passing through the squares).

\subsection{The region \protect\ref{N3R2}
of small
$\boldsymbol{|U_{\boldsymbol{e1}}|^2}$
and large
$\boldsymbol{|U_{\boldsymbol{\mu1}}|^2}$}
\label{SN3R2}

From the discussion in Section \ref{SN3R1}
it follows that,
if the indications in favor of
$ \nu_{\mu} \leftrightarrows \nu_{e} $
oscillations
reported by the LSND collaboration
\cite{LSND}
are confirmed by future experiments,
the parameters
$ \left| U_{e1} \right|^2 $
and
$ \left| U_{\mu1} \right|^2 $
must satisfy the inequalities
\begin{equation}
\left| U_{e1} \right|^2
\le
a_{e}^{0}
\qquad \mbox{and} \qquad
\left| U_{\mu1} \right|^2
\ge
1 - a_{\mu}^{0}
\;.
\label{229}
\end{equation}
with the quantities
$a_{e}^{0}$ and $a_{\mu}^{0}$
given by Eq.(\ref{208}).

Taking into account the unitarity of the mixing matrix
and
Eq.(\ref{229}) we have
\begin{equation}
\left| U_{e1} \right|^2
\leq
1 - \left| U_{\mu1} \right|^2
\leq
a_{\mu}^{0}
\;.
\label{230}
\end{equation}
From Eqs.(\ref{229}) and (\ref{230})
we have
\begin{equation}
\left| U_{e1} \right|^2
\leq
\mbox{Min}\left[ a_{e}^{0} , a_{\mu}^{0} \right]
\;.
\label{231}
\end{equation}
The inequalities (\ref{229})
and the unitarity of the mixing matrix
imply that
also
$ \left| U_{\tau1} \right|^2 $
is small:
\begin{equation}
\left| U_{\tau1} \right|^2
\leq
1 - \left| U_{\mu1} \right|^2
\leq
a_{\mu}^{0}
\;.
\label{232}
\end{equation}
Taking into account the inequalities
(\ref{229})--(\ref{232}),
from Eq.(\ref{203})
it follows that the oscillation amplitude
$ A_{\nu_{e};\nu_{\tau}} $
is quadratic in the small quantities
$ \left| U_{e1} \right|^2 $
and
$ \left| U_{\tau1} \right|^2 $,
whereas the oscillation amplitudes
$ A_{\nu_{\mu};\nu_{e}} $
and
$ A_{\nu_{\mu};\nu_{\tau}} $
are linear in the same small quantities.
For the oscillation amplitude
$ A_{\nu_{e};\nu_{\tau}} $
we have the following upper bound:
\begin{equation}
A_{\nu_{e};\nu_{\tau}}
\le
4
\,
\mbox{Min}\left[ a_{e}^{0} , a_{\mu}^{0} \right]
\,
a_{\mu}^{0}
\;.
\label{233}
\end{equation}
In Fig.\ref{n3r2elta}
we have plotted the upper bound
for the oscillation amplitude
$ A_{\nu_{e};\nu_{\tau}} $
obtained
from Eq.(\ref{233}) using the results
of the Bugey, CDHS and CCFR84 experiments
(the curve passing through the triangles).
The solid line in Fig.\ref{n3r2elta}
is the exclusion curve
obtained in the FNAL E531 experiment.
Taking into account the unitarity constraint
$ A_{\nu_{e};\nu_{\tau}} \leq B_{\nu_{e};\nu_{e}} $,
in Fig.\ref{n3r2elta}
we have also plotted the exclusion curve
obtained in the Bugey experiment.
From Fig.\ref{n3r2elta} it can be seen that
in all the considered range of $ \Delta m^2 $
the upper bound for $ A_{\nu_{e};\nu_{\tau}} $
given by Eq.(\ref{233})
is very small,
varying from about $10^{-3}$ to about $10^{-2}$
for
$ \Delta m^2 \gtrsim 0.3 \, \mbox{eV}^2 $.

Additional limits
on the oscillation amplitude
$ A_{\nu_{e};\nu_{\tau}} $
can be obtained by
taking into account the results of
the experiments on the search for
$ \nu_\mu \leftrightarrows \nu_e $
and
$ \nu_\mu \leftrightarrows \nu_\tau $
oscillations.
In the linear approximation in
the small quantities
$|U_{e1}|^2$, $(1-|U_{\mu1}|^2)$ and $|U_{\tau1}|^2$,
from Eq.(\ref{203}) we have
\arraycolsep=0pt
\begin{eqnarray}
&&
|U_{e1}|^2
\simeq
{\displaystyle
A_{\nu_{\mu};\nu_{e}}
\over\displaystyle
4
}
\;,
\label{234}
\\
&&
|U_{\tau1}|^2
\simeq
{\displaystyle
A_{\nu_{\mu};\nu_{\tau}}
\over\displaystyle
4
}
\;.
\label{235}
\end{eqnarray}
With the help of Eqs.(\ref{232}) and (\ref{234})
we obtain the following upper bound
for the oscillation amplitude
$ A_{\nu_{e};\nu_{\tau}} $:
\begin{equation}
A_{\nu_{e};\nu_{\tau}}
\le
A_{\nu_{\mu};\nu_{e}}^{0}
\,
a_{\mu}^{0}
\;.
\label{236}
\end{equation}
In Fig.\ref{n3r2elta}
we have plotted
the corresponding boundary curve
obtained from the results
of the CDHS, CCFR84 and BNL E776 experiments
(the curve passing through the squares).
From this figure it can be seen
that for
$ \Delta m^2 \gtrsim 1 \, \mbox{eV}^2 $
the upper bound on the oscillation amplitude
$ A_{\nu_{e};\nu_{\tau}} $
varies from $ 2 \times 10^{-5} $ to $ 5 \times 10^{-4} $
and
is more stringent than that obtained
from the results of disappearance experiments
using Eq.(\ref{233}).

Finally,
from Eqs.(\ref{234}) and (\ref{235})
we have
\begin{equation}
A_{\nu_{e};\nu_{\tau}}
\lesssim
{\displaystyle
A_{\nu_{\mu};\nu_{e}}^{0}
\,
A_{\nu_{\mu};\nu_{\tau}}^{0}
\over\displaystyle
4
}
\label{237}
\end{equation}
The corresponding boundary curve obtained
from the results
of the BNL E776, FNAL E531 and CCFR95 experiments
is presented in Fig.\ref{n3r2elta}
(the curve passing through the circles).
From this figure it can be seen that for
$ \Delta m^2 \gtrsim 4 \, \mbox{eV}^2 $
the amplitude of
$ \nu_e \leftrightarrows \nu_\tau $
oscillations
is smaller than $ 3 \times 10^{-5} $.

Thus,
if the parameter
$|U_{e1}|^2$ is small and
the parameter $|U_{\mu1}|^2$ is large,
$ \nu_{e} \leftrightarrows \nu_{\tau} $
oscillations are strongly suppressed.
On the other hand, there are no constraints
on the amplitudes of
$ \nu_{\mu} \leftrightarrows \nu_{e} $
and
$ \nu_{\mu} \leftrightarrows \nu_{\tau} $
oscillations.
As is well known,
new experiments
on the search for
$ \nu_\mu \to \nu_\tau $
transitions
are under way at CERN
(CHORUS
\cite{CHORUS}
and NOMAD
\cite{NOMAD}).
Experiments
on the search for
$ \nu_\mu \to \nu_e $
transitions
are continuing at Los Alamos
(LSND \cite{LSND})
and at Rutherford Appleton Laboratory
(KARMEN \cite{KARMEN}).

Let us conclude this Section
with the following remarks.
In the general case of mixing,
the flavor neutrinos
$\nu_e$,
$\nu_\mu$ and
$\nu_\tau$
are not particles with a definite mass.
Effective masses of the flavor neutrinos
can be introduced in the framework
of specific models.
This is the case of the model under
consideration.
If the mixing parameters are in
the region \ref{N3R1}
$\nu_e$ and $\nu_\mu$ are heavy
and
$\nu_\tau$ is the lightest neutrino:
$ m_{\nu_e} \simeq m_3 $,
$ m_{\nu_\mu} \simeq m_3 $ and
$ m_{\nu_\tau} \simeq m_1 $.
If the mixing parameters are in
the region \ref{N3R2}
$\nu_e$ and $\nu_\tau$ are heavy
and
$\nu_\mu$ is the lightest neutrino:
$ m_{\nu_e} \simeq m_3 $,
$ m_{\nu_\tau} \simeq m_3 $ and
$ m_{\nu_\mu} \simeq m_1 $.
Such inverted neutrino mass spectra
are compatible with
the r-process production of heavy elements
in the neutrino-heated ejecta of supernovae
\cite{r-process}.

\section{Mixing of four massive neutrinos}
\label{SN4}

All the existing
indications in favor of
neutrino masses and mixing
(solar neutrinos, atmospheric neutrinos,
dark matter, LSND)
require at least three different scales of
squared-mass differences.
In this section we will consider
short-baseline oscillations of
the terrestrial neutrinos
in a scheme with mixing of four massive neutrinos.
We will consider a specific scheme
\cite{CM93,PV93,SSF93,PHKC95,GCGG95,SG95}
with two light neutrinos $\nu_1$, $\nu_2$
and two neutrinos
$\nu_3$, $\nu_4$ with masses in the eV range.
We will assume that the value of
$ \Delta m^2_{43} \equiv m^2_4 - m^2_3 $
is relevant for the suppression of the flux
of solar $\nu_e$'s
and the value of
$ \Delta m^2_{21} \equiv m^2_2 - m^2_1 $
could be relevant for the explanation
of the atmospheric neutrino anomaly.
Thus, we have
\begin{equation}
\Delta m^2_{43}
\ll
\Delta m^2_{21}
\ll
\Delta m^2
\;,
\label{301}
\end{equation}
with
$ \Delta m^2 \equiv m^2_4 - m^2_1 $.

The important difference between the model considered here
and that considered in Section \ref{SN3}
is that in this model
transitions of active neutrinos
into sterile states are possible.
The transition probabilities between
different states
are given by the general expressions
(\ref{110}) and (\ref{112})
with the following oscillation amplitudes:
\arraycolsep=0pt
\begin{eqnarray}
&&
A_{\nu_{\alpha};\nu_{\beta}}
=
4
\left|
\sum_{j=1,2}
U_{{\beta}j}
U_{{\alpha}j}^{*}
\right|^2
\;,
\label{302}
\\
&&
B_{\nu_{\alpha};\nu_{\alpha}}
=
4
\,
b_{\alpha}
\left(
1
-
b_{\alpha}
\right)
\;,
\label{303}
\end{eqnarray}
where
\begin{equation}
b_{\alpha}
\equiv
\sum_{j=1,2}
\left| U_{{\alpha}j} \right|^2
\label{304}
\end{equation}

As in Section \ref{SN3},
we start with the implications
in the scheme under consideration
of the results
of the reactor and accelerator
disappearance experiments.
It is clear that we can use
the results presented in Section \ref{SN3}.
We have
\arraycolsep=0pt
\begin{eqnarray}
&&
b_{\alpha}
\le
a_{\alpha}^{0}
\label{305}
\\
\mbox{or}
&&
\nonumber
\\
&&
b_{\alpha}
\ge
1 - a_{\alpha}^{0}
\;,
\label{306}
\end{eqnarray}
with $\alpha=e,\mu$.
The quantities $a_{\alpha}^{0}$
are determined by Eq.(\ref{208}).
For each value of
$ \Delta m^2 $
we determined the value of
$a_{e}^{0}$ and
$a_{\mu}^{0}$
from the exclusion plots
obtained in the Bugey \cite{Bugey95},
CDHS \cite{CDHS84} and CCFR84 \cite{CCFR84}
experiments.
We will consider values of
$ \Delta m^2 $
in the interval
$ 0.5 \, \mathrm{eV}^2 \lesssim
\Delta m^2
\lesssim 10^{2} \, \mathrm{eV}^2 $,
where
the quantities
$a_{e}^{0}$ and
$a_{\mu}^{0}$
are smaller than 0.1.

Thus,
the values of the parameters
$ b_{e} $
and
$ b_{\mu} $
can lie in one of the following four regions:

\def\theenumi{\Roman{enumi}}

\begin{enumerate}

\item \label{N4R1}
The region of small
$ b_{e} $
and
$ b_{\mu} $.

\item \label{N4R2}
The region of small
$ b_{e} $
and large
$ b_{\mu} $.

\item \label{N4R3}
The region of large
$ b_{e} $
and small
$ b_{\mu} $.

\item \label{N4R4}
The region of large
$ b_{e} $
and
$ b_{\mu} $.

\end{enumerate}

If the neutrino masses satisfy
the relation (\ref{301})
and
$ \Delta m^2_{43} $
is relevant for the suppression of the
solar $\nu_e$ flux,
the probability of solar $\nu_e$'s
to survive is given by
(see the Appendix \ref{APA})
\begin{equation}
P_{\nu_e\to\nu_e}
=
\left(
1
-
\sum_{j=1,2}
\left| U_{ej} \right|^2
\right)^2
P_{\nu_e\to\nu_e}^{(3,4)}
+
\sum_{j=1,2}
\left| U_{ej} \right|^4
\;,
\label{307}
\end{equation}
where
$ P_{\nu_e\to\nu_e}^{(3,4)} $
is the survival probability
due to the mixing of $\nu_e$
with $\nu_3$ and $\nu_4$.
If the parameter
$b_{e}$
is large,
for the survival probability we have
\begin{equation}
P_{\nu_e\to\nu_e}
\simeq
\sum_{j=1,2}
\left| U_{ej} \right|^4
\;.
\label{308}
\end{equation}
Thus,
if the parameters $b_{e}$ and $b_{\mu}$
lie in the region \ref{N4R3} or in the region \ref{N4R4},
the probability
$ P_{\nu_e\to\nu_e} $
practically
does not depend on the neutrino energy.
Moreover,
taking into account that
in the regions \ref{N4R3} and \ref{N4R4}
we have
$ \displaystyle
\sum_{j=1,2}
\left| U_{ej} \right|^2 \simeq 1
$,
from Eq.(\ref{308})
we obtain
\begin{equation}
P_{\nu_e\to\nu_e}
\gtrsim
{ 1 \over 2 }
\;.
\label{309}
\end{equation}
The case of a constant
$ P_{\nu_e\to\nu_e} $
is disfavored by the solar neutrino data.
In Ref.\cite{KP96}
it was shown that
the solar neutrino data
cannot be explained with a constant
$ P_{\nu_e\to\nu_e} $
if the fluxes of
$^8\mbox{B}$ and  $^7\mbox{Be}$ neutrinos
are allowed to vary in rather wide intervals
around the standard model values.

In both regions \ref{N4R1} and \ref{N4R2}
we have
$ \displaystyle
\sum_{k=3,4}
\left| U_{ek} \right|^2
\simeq 1
$.
Thus,
information about the value of the masses
$ m_3 \simeq m_4 $
can be obtained from the investigation
of the end-point part of the $\beta$-spectrum
of $^3$H decay
(see Eq.(\ref{211})).
If massive neutrinos are Majorana particles,
neutrinoless double-beta decay is possible.
It is clear that all the relations
referring to this process
derived in Section \ref{SN3}
in the framework of the scheme
with three massive neutrinos
(see Eqs.(\ref{213})--(\ref{223}))
are also valid in the scheme
under consideration.

In the following we will discuss
the implications of the results
of neutrino oscillations appearance experiments
if the values of the parameters
$b_{e}$ and $b_{\mu}$
lie in the region \ref{N4R1} or
in the region \ref{N4R2}.

\subsection{The region \protect\ref{N4R1}
of small
$\boldsymbol{b_{\boldsymbol{e}}}$
and
$\boldsymbol{b_{\boldsymbol{\mu}}}$}
\label{SN4R1}

In this region
$ \nu_{\mu} \leftrightarrows \nu_{e} $
oscillations are suppressed.
In fact,
using the Cauchy-Schwarz inequality,
from Eqs.(\ref{302}) and (\ref{304})
we obtain
\begin{equation}
A_{\nu_{\mu};\nu_{e}}
\leq
4 \, b_{\mu} \, b_{e}
\;.
\label{310}
\end{equation}
Taking into account that in this region
$ b_{\mu} \le a^{0}_{\mu} $
and
$ b_{e} \le a^{0}_{e} $,
we obtain
\begin{equation}
A_{\nu_{\mu};\nu_{e}}
\leq
4 \, a^{0}_{\mu} \, a^{0}_{e}
\;.
\label{311}
\end{equation}
In Fig.\ref{n4r1muel}
we have plotted the curve
that represent this
upper bound obtained
from the results
of the Bugey,
CDHS and CCFR84 experiments
(the curve passing through the circles).

Let us consider now the
neutrino oscillation appearance
experiments.
From the unitarity relation (\ref{113}) we have
\arraycolsep=0pt
\begin{eqnarray}
&&
A_{\nu_{\mu};\nu_{\tau}}
\leq
B_{\nu_{\mu};\nu_{\mu}}
\;,
\label{312}
\\
&&
A_{\nu_{e};\nu_{\tau}}
\leq
B_{\nu_{e};\nu_{e}}
\;.
\label{313}
\end{eqnarray}
In the case of three massive neutrinos
considered in Section \ref{SN3}
instead of these inequalities we had
two approximate equalities,
that allowed us to obtain additional
strong limitations for the amplitude of
$ \nu_{\mu} \leftrightarrows \nu_{e} $
oscillations
(see Eq.(\ref{228})).
In the model under consideration
transitions of $\nu_\mu$ and $\nu_e$
into sterile states are possible and,
instead of approximate equalities
such as Eq.(\ref{227}),
we have the inequalities
\arraycolsep=0pt
\begin{eqnarray}
&&
b_{\mu}
\gtrsim
{ A_{\nu_{\mu};\nu_{\tau}} \over 4 }
\;,
\label{314}
\\
&&
b_{e}
\gtrsim
{ A_{\nu_{e};\nu_{\tau}} \over 4 }
\;.
\label{315}
\end{eqnarray}
The inequality (\ref{314})
can be useful
if a positive result is found
in the experiments on the search for
$ \nu_\mu \leftrightarrows \nu_\tau $
oscillations
(CHORUS
\cite{CHORUS},
NOMAD
\cite{NOMAD},
COSMOS
\cite{COSMOS}).

An important difference between the model considered here
and the one considered in Section \ref{SN3}
is that,
in the case of mixing of four massive neutrinos,
in the region \ref{N4R1}
of small $b_{e}$ and $b_{\mu}$
the result of the LSND
experiment is compatible
with the results of all the other
neutrino oscillation experiments
for
$ 5 \, \mathrm{eV}^2 \lesssim
\Delta m^2
\lesssim 70 \, \mathrm{eV}^2 $.

\subsection{The region \protect\ref{N4R2}
of small
$\boldsymbol{b_{\boldsymbol{e}}}$
and large
$\boldsymbol{b_{\boldsymbol{\mu}}}$}
\label{SN4R2}

We will consider now the region \ref{N4R2}
of small $b_{e}$ and large $b_{\mu}$.
From Eq.(\ref{303})
in the linear approximation in the small quantities
$b_{e}$ and $1-b_{\mu}$
we have
\arraycolsep=0pt
\begin{eqnarray}
&&
b_{e}
\simeq
{ B_{\nu_{e};\nu_{e}} \over 4 }
\;,
\label{317}
\\
&&
1 - b_{\mu}
\simeq
{ B_{\nu_{\mu};\nu_{\mu}} \over 4 }
\;.
\label{318}
\end{eqnarray}
The essential difference of
the model under consideration
from the model with mixing of three neutrinos
considered in the previous section
is that,
in the case of mixing of four massive neutrinos,
$ \nu_e \leftrightarrows \nu_\tau $
oscillations could be not suppressed
in the region \ref{N4R2}.
This is due to the fact that
in the model under consideration
transitions of active neutrinos
into sterile states are possible.
In fact,
using the Cauchy-Schwarz inequality,
from Eq.(\ref{302})
we have
\begin{equation}
A_{\nu_{e};\nu_{\tau}}
\leq
4 \, b_{e} \, b_{\tau}
\;.
\label{319}
\end{equation}
Furthermore,
from the unitarity of the mixing matrix it follows that
\begin{equation}
\sum_{\alpha} b_{\alpha} = 2
\;.
\label{320}
\end{equation}
From this equation we cannot conclude
that in the region of small
$b_{e}$ and $1-b_{\mu}$
the parameter $b_{\tau}$ is small.
Thus,
the right-hand side of Eq.(\ref{319})
could be not quadratic in small quantities
(as it is in the case of
mixing of three  massive neutrinos)
and
$A_{\nu_{e};\nu_{\tau}}$
could be not suppressed.

Let us notice that from the unitarity relation (\ref{113})
and from Eqs.(\ref{317}) and (\ref{318})
we have the following lower bounds for
$b_{e}$ and $1-b_{\mu}$
\arraycolsep=0pt
\begin{eqnarray}
&&
b_{e}
\gtrsim
{ A_{\nu_{\mu};\nu_{e}} \over 4 }
\;,
\label{321}
\\
&&
1-b_{\mu}
\gtrsim
{ A_{\nu_{\mu};\nu_{e}} + A_{\nu_{\mu};\nu_{\tau}} \over 4 }
\;.
\label{322}
\end{eqnarray}

\subsection{Atmospheric neutrinos}

Up to now we have considered
the constraints on the value of the mixing
parameters
$b_{e}$ and $b_{\mu}$
due to the results of the solar neutrino experiments
and of the
reactor and accelerator neutrino oscillation experiments.
If the squared-mass difference
$ \Delta m^2_{21} $
is relevant for the oscillation of atmospheric neutrinos
(which is natural in this scheme),
we can obtain an additional constraint
on
$b_{e}$ and $b_{\mu}$
from the results of the
experiments on the detection of
atmospheric neutrinos.

In the model under consideration
the survival probability
of atmospheric $\nu_{\alpha}$
is given by
(see the Appendix \ref{APB})
\begin{equation}
P_{\nu_{\alpha}\to\nu_{\alpha}}
=
\left(
1
-
\sum_{k=3,4}
\left| U_{{\alpha}k} \right|^2
\right)^2
P_{\nu_{\alpha}\to\nu_{\alpha}}^{(1,2)}
+
\left(
\sum_{k=3,4}
\left| U_{{\alpha}k} \right|^2
\right)^2
\;,
\label{324}
\end{equation}
where
$ P_{\nu_{\alpha}\to\nu_{\alpha}}^{(1,2)} $
is the survival probability
due to the mixing of $\nu_{\alpha}$
with $\nu_1$ and $\nu_2$.
In the region \ref{N4R1},
where both $b_{e}$ and $b_{\mu}$
are small,
we have
\arraycolsep=0pt
\begin{eqnarray}
&&
P_{\nu_{\mu}\to\nu_{\mu}}
=
b_{\mu}^2
\,
P_{\nu_{\mu}\to\nu_{\mu}}^{(1,2)}
+
\left( 1 - b_{\mu} \right)^2
\simeq
1
\;,
\label{325}
\\
&&
P_{\nu_{e}\to\nu_{e}}
=
b_{e}^2
\,
P_{\nu_{e}\to\nu_{e}}^{(1,2)}
+
\left( 1 - b_{e} \right)^2
\simeq
1
\;.
\label{326}
\end{eqnarray}
Thus,
if the parameters
$b_{e}$ and $b_{\mu}$
are in the region \ref{N4R1},
the ratio
of the observed
neutrino-induced $e$-like and $\mu$-like
events
in the atmospheric neutrino detectors
should be equal to the expected ratio.
This is in contradiction with the
results of the
Kamiokande
\cite{Kamiokande-atmospheric},
IMB
\cite{IMB}
and Soudan
\cite{Soudan}
experiments.
On the other hand,
if the parameters
$b_{e}$ and $b_{\mu}$
are in the region \ref{N4R2},
from Eq.(\ref{324})
we obtain
\arraycolsep=0pt
\begin{eqnarray}
&&
P_{\nu_{\mu}\to\nu_{\mu}}
\simeq
P_{\nu_{\mu}\to\nu_{\mu}}^{(1,2)}
\;,
\label{327}
\\
&&
P_{\nu_{e}\to\nu_{e}}
\simeq
1
\;,
\label{328}
\end{eqnarray}
and the results of the
Kamiokande, IMB and Soudan experiments
can be explained with
$ \nu_\mu \leftrightarrows \nu_\tau $
oscillations.

Therefore,
if the atmospheric neutrino anomaly
found in the
Kamiokande, IMB and Soudan experiments
is confirmed by future experiments,
the values of the mixing parameters
$b_{e}$ and $b_{\mu}$
can lie only in the region \ref{N4R2}.

\section{Conclusions}
\label{Conclusions}

We have considered short-baseline neutrino oscillations,
$^3$H $\beta$-decay
and
$(\beta\beta)_{0\nu}$-decay
in two specific schemes
with mixing of three and four
massive neutrino fields,
respectively.

Having in mind the experimental indications
in favor of neutrino mass and mixing,
we have assumed that
in the case of three neutrinos
the neutrino mass spectrum is composed
of one very light neutrino
with mass
$ m_1 \ll 1 \, \mbox{eV} $
and two neutrinos with masses
$ m_2 \simeq m_3 $ in the eV range
and a squared-mass difference
$ m_3^2 - m_2^2 $
which is
relevant for the suppression
of the flux of solar $\nu_e$'s.
In this scheme,
the oscillations of the terrestrial neutrinos
are determined by three parameters,
$ \Delta m^2 \equiv m_3^2 - m_1^2 $
and the two mixing parameters
$|U_{e1}|^2$
and
$|U_{\mu1}|^2$.
From the analysis of the results of
reactor and accelerator
disappearance experiments
and from the results of the solar neutrino experiments,
it follows that only two regions of the values
of the mixing parameters are allowed:
\ref{N3R1}.
$|U_{e1}|^2$
and
$|U_{\mu1}|^2$
are both small;
\ref{N3R2}.
$|U_{e1}|^2$
is small
and
$|U_{\mu1}|^2$
is large (close to one).
We have shown that in the region \ref{N3R1}
the oscillations
$ \nu_\mu \leftrightarrows \nu_e $
are suppressed
and the result of the LSND experiment
is not compatible with
the results of all the other neutrino
oscillation experiments.
Instead,
in the region \ref{N3R2}
the oscillations
$ \nu_e \leftrightarrows \nu_\tau $
are suppressed
and we have obtained rather strong limits
for the oscillation amplitude
$ A_{\nu_{e};\nu_{\tau}} $
from the results of neutrino oscillation experiments.
If the masses $m_2$ and $m_3$
are in the eV region,
the effect of neutrino masses
could be observed in the next generation of experiments
on the measurement of the end-point part
of the $^3$H $\beta$-decay spectrum
and in the next generation of experiments
on the search for
$(\beta\beta)_{0\nu}$-decay.
We have shown that,
if these experiment find a positive effect,
a comparison of the results of the
$(\beta\beta)_{0\nu}$-decay experiments
and of the $^3$H $\beta$-decay experiments
could allow to obtain direct information
about the CP violation in the lepton sector.

In the case of mixing of four massive neutrino fields
we have assumed that the spectrum of neutrino masses
is composed of two very light masses
$ m_1 \simeq m_2 \ll 1 \, \mbox{eV} $
with a value of $ m_2^2 - m_1^2 $ which could be
relevant for the explanation of the atmospheric
neutrino anomaly
and two masses $ m_3 \simeq m_4 $ in the eV range
with a value of $ m_4^2 - m_3^2 $ which is
relevant for the suppression of the flux of solar $\nu_e$'s.
Since in this model
transitions from active to sterile states are possible,
the unitarity of the mixing matrix
does not put as strong constraints
on the oscillation channels
as in the case of three neutrinos.
The survival probabilities of
$\nu_e$ and $\nu_\mu$
are determined by the two parameters
$ \displaystyle
b_{\alpha}
=
\sum_{j=1,2}
|U_{{\alpha}j}|^2
$
(with $\alpha=e,\mu$).
From the results of
neutrino oscillation disappearance experiments
and the results of solar neutrino experiments
it follows that
only two regions of values of the parameters
$b_{e}$ and $b_{\mu}$ are allowed:
\ref{N4R1}.
The region of small
$ b_{e} $
and
$ b_{\mu} $;
\ref{N4R2}.
The region of small
$ b_{e} $
and large
$ b_{\mu} $.
Oscillations
$ \nu_\mu \leftrightarrows \nu_e $
are suppressed in the region \ref{N4R1},
but the LSND result is compatible with
the results of other neutrino oscillation experiments
if 
$ \Delta m^2 = m^2_4 - m^2_1 $
is in the range
$ 5 \, \mathrm{eV}^2 \lesssim
\Delta m^2
\lesssim 70 \, \mathrm{eV}^2 $.
If the parameters
$ b_{e} $
and
$ b_{\mu} $
are in the region \ref{N4R2},
$ \nu_e \leftrightarrows \nu_\tau $
oscillations
could be not suppressed
(unlike the case of three neutrinos).
If the atmospheric neutrino anomaly
is confirmed by future experiments,
the region \ref{N4R1} will be excluded.

Both schemes considered here
are in agreement with the
cold and hot dark matter scenario
with two practically degenerate neutrinos
with masses in the eV range
\cite{PHKC95}.

From the results of
neutrino oscillation disappearance experiments
and solar neutrino experiments
it follows that
in both schemes considered here
the electron neutrino is "heavy".
This means that both schemes
are compatible with the constraints
coming from the r-process nucleosynthesis
in the neutrino-heated ejecta of supernovae
\cite{r-process}.

In conclusion,
let us remark that,
if one of the neutrino mixing schemes
considered here
is realized in nature,
i.e. there is no hierarchy of neutrino masses,
the neutrino mass spectrum
has a completely different character from the
mass spectra of quarks and charged leptons.
From our analysis of the results of
neutrino oscillation experiments
it follows that
there is no hierarchy of couplings
among generations in the lepton sector,
i.e.
the mixing matrix of neutrinos
is completely different from that of quarks.

\acknowledgments

S.B. would like to acknowledge
the kind hospitality of the
Elementary Particle Sector of SISSA,
where this work has been done.
C.W.K. wishes to thank
the Department of Physics,
Korea Advanced Institute of Science and Technology,
for the hospitality extended to him
while this work was in progress.

\appendix

\section{Survival probability of solar neutrinos}
\label{APA}

In this appendix we will derive the formula (\ref{307})
for the survival probability
$ P_{\nu_{e}\to\nu_{e}} $
of solar neutrinos
in the case of mixing of four massive neutrino fields
considered in Section \ref{SN4}.

In general,
the survival probability of solar $\nu_e$'s
can be written as
\begin{equation}
P_{\nu_{e}\to\nu_{e}}
=
\left|
\sum_{jk}
U_{ej}^{(M)}
\,
\mathcal{A}_{jk}
\,
U_{ek}^{*}
\right|^2
\;,
\label{551}
\end{equation}
where
$ U^{(M)} $
is the mixing matrix in matter in the point of 
$\nu_e$ production in the Sun
and
$ \mathcal{A}_{jk} $
is the amplitude
of
$ \nu^{\mathrm{m}}_{j} \to \nu_{k} $
transitions
when the solar neutrinos travel from the central part
of the Sun to the Earth,
$\nu^{\mathrm{m}}_{j}$
being the effective mass eigenstate neutrinos in matter
in the point of
$\nu_e$ production in the Sun
(see, e.g., \cite{BP87,CWKim}). 
Equation (\ref{551})
is valid both for vacuum oscillations
and for resonant MSW transitions
of solar $\nu_e$'s.
In the case of vacuum oscillations
$ U^{(M)} = U $,
$ \nu^{\mathrm{m}}_{j} = \nu_{j} $
and
\begin{equation}
\mathcal{A}_{jk}
=
\delta_{jk}
\exp
\left[
- i
\int_{x_i}^{x_f}
E_j \, \mbox{d}x
\right]
\;,
\label{552}
\end{equation}
where
$ E_j \simeq p + m_j^2 / 2 p $,
and
$x_i$ and $x_f$
are the initial and final points
of the neutrino trajectory.

In the model under consideration
it is assumed that
$ \Delta m^2_{43} $
is relevant for the suppression of the flux
of solar $\nu_e$'s
and
$
\Delta m^2_{41}
\gg
\Delta m^2_{21}
\gg
\Delta m^2_{43}
$.
This implies that,
if $\Delta m^2_{43}$
is in the range of the MSW solution
of the solar neutrino problem,
the transitions
between
$\nu^{\mathrm{m}}_{4(3)}$
and
$\nu_{3(4)}$ in the Sun can be strongly
affected by solar matter effects.
At the same time,
the resonance densities associated with
$
\Delta m^2_{41}
\simeq
\Delta m^2_{42}
\simeq
\Delta m^2_{31}
\simeq
\Delta m^2_{32}
$
and with
$\Delta m^2_{21}$
are much bigger than the density in the Sun
and therefore the evolution
of the $\nu_1$ and $\nu_2$ states
in the Sun is adiabatic  
(see the discussions in Refs.\cite{KP88,KP86}).
As a consequence 
we have
\begin{equation}
U_{ej}^{(M)} = U_{ej}
\qquad \mbox{for} \qquad
j=1,2
\;,
\label{553}
\end{equation}
and
Eq.(\ref{552})
is valid for $j,k=1,2$
and for $j=1,2$ and $k=3,4$.
Thus,
the survival probability
averaged over the neutrino energy spectrum
and the region of neutrino production in the Sun
can be written as
\begin{equation}
P_{\nu_{e}\to\nu_{e}}
=
\sum_{j=1,2}
\left| U_{ej} \right|^4
+
\left|
\sum_{k,k'=3,4}
U_{ek}^{(M)}
\,
\mathcal{A}_{kk'}
\,
U_{ek'}^{*}
\right|^2
\;.
\label{554}
\end{equation}
Using the unitarity of the mixing matrix,
we can write Eq.(\ref{554}) as
\begin{equation}
P_{\nu_e\to\nu_e}
=
\sum_{j=1,2}
\left| U_{ej} \right|^4
+
\left(
1
-
\sum_{j=1,2}
\left| U_{ej} \right|^2
\right)^2
P_{\nu_e\to\nu_e}^{(3,4)}
\;,
\label{555}
\end{equation}
where
\begin{equation}
P_{\nu_e\to\nu_e}^{(3,4)}
=
\left|
\sum_{k,k'=3,4}
{\displaystyle
U_{ek}^{(M)}
\over\displaystyle
\sqrt{
\sum_{k''=3,4}
\left| U_{ek''}^{(M)} \right|^2
}
}
\,
\mathcal{A}_{kk'}
\,
{\displaystyle
U_{ek'}^{*}
\over\displaystyle
\sqrt{
\sum_{k''=3,4}
\left| U_{ek''} \right|^2
}
}
\right|^2
\;.
\label{556}
\end{equation}

In the case of vacuum oscillations
Eq.(\ref{556}) can be written as
\begin{equation}
P_{\nu_{\alpha}\to\nu_{\alpha}}^{(3,4)}
=
1 - {1\over2}
\sin^2 2\theta
\left(
1
-
\cos
{\displaystyle
\Delta m^2_{43} \, L
\over\displaystyle
2 \, p
}
\right)
\;,
\label{557}
\end{equation}
which is the standard two-generation formula.
Here
the mixing angle
$\theta$
is defined by
\begin{equation}
\cos \theta
=
{\displaystyle
|U_{e3}|
\over\displaystyle
\sqrt{
\left(
\sum_{k=3,4}
\left| U_{ek} \right|^2
\right)^2
}
}
\;,
\null \hskip2cm \null \displaystyle
\sin \theta
=
{\displaystyle
|U_{e4}|
\over\displaystyle
\sqrt{
\left(
\sum_{k=3,4}
\left| U_{ek} \right|^2
\right)^2
}
}
\;.
\label{558}
\end{equation}

In the case of MSW resonant transitions
Eq.(\ref{556}) can be written as
\begin{equation}
P_{\nu_e\to\nu_e}^{(3,4)}
=
{1\over2}
+
\left(
{1\over2}
-
P_{34}
\right)
\cos 2\theta_M
\cos 2\theta
\;,
\label{559}
\end{equation}
which has the form of the standard two-generation formula.
Here
$ P_{34} = \left| \mathcal{A}_{34} \right|^2$
is the probability of
$ \nu^{\mathrm{m}}_3 \to \nu_4 $
transitions,
which is equal to
the probability of
$ \nu^{\mathrm{m}}_4 \to \nu_3 $
transitions.
The mixing angle
$\theta$ in vacuum
is defined by Eq.(\ref{558})
and
the mixing angle
$\theta_M$ in matter
is defined by
\begin{equation}
\cos \theta_M
=
{\displaystyle
\left| U_{e3}^{(M)} \right|
\over\displaystyle
\sqrt{
\left(
\sum_{k=3,4}
\left| U_{ek}^{(M)} \right|^2
\right)^2
}
}
\;,
\null \hskip2cm \null \displaystyle
\sin \theta_M
=
{\displaystyle
\left| U_{e4}^{(M)} \right|
\over\displaystyle
\sqrt{
\left(
\sum_{k=3,4}
\left| U_{ek}^{(M)} \right|^2
\right)^2
}
}
\;.
\label{560}
\end{equation}

\section{Survival probability of atmospheric neutrinos}
\label{APB}

In this appendix we will derive the formula (\ref{324})
for the survival probability
$ P_{\nu_{\alpha}\to\nu_{\alpha}} $
of atmospheric neutrinos
in the case of mixing of four massive neutrino fields
considered in Section \ref{SN4}.

In general,
the vacuum oscillation survival probability
of $\nu_{\alpha}$ is given by
(see, for example, Refs.\cite{BP78,BP87,CWKim})
\begin{equation}
P_{\nu_{\alpha}\to\nu_{\alpha}}
=
\sum_{j,k}
|U_{{\alpha}j}|^2
\,
|U_{{\alpha}k}|^2
\,
\exp
\left(
- i \,
{\displaystyle
\Delta m^2_{jk} L
\over\displaystyle
2 p
}
\right)
\;,
\label{501}
\end{equation}
with
$ \Delta m^2_{jk} \equiv m^2_j - m^2_k $.

In the model under consideration,
for the atmospheric neutrinos we have
\begin{equation}
\exp
\left(
- i \,
{\displaystyle
\Delta m^2_{43} L
\over\displaystyle
2 p
}
\right)
\simeq 1
\;.
\label{502}
\end{equation}
After averaging over the neutrino energy
and the distance between the points
of neutrino production and detection,
we have
\begin{equation}
\left\langle
\exp
\left(
- i \,
{\displaystyle
\Delta m^2_{jk} L
\over\displaystyle
2 p
}
\right)
\right\rangle
\simeq 0
\qquad \mbox{for} \qquad
j=1,2
\quad \mbox{and} \quad
k=3,4
\;.
\label{503}
\end{equation}

From Eqs.(\ref{501})--(\ref{503}),
for the survival probability
of $nu_{\alpha}$ we obtain
\begin{equation}
P_{\nu_{\alpha}\to\nu_{\alpha}}
=
\left(
\sum_{k=3,4}
\left| U_{{\alpha}k} \right|^2
\right)^2
+
\sum_{j=1,2}
\left| U_{{\alpha}j} \right|^4
+
2 \, |U_{\alpha1}|^2 \, |U_{\alpha2}|^2 \,
\cos
{\displaystyle
\Delta m^2_{21} \, L
\over\displaystyle
2 \, p
}
\label{504}
\end{equation}

Using the unitarity relation (\ref{108})
we have
\begin{equation}
\sum_{j=1,2}
\left| U_{{\alpha}j} \right|^4
=
\left(
1
-
\sum_{k=3,4}
\left| U_{{\alpha}k} \right|^2
\right)^2
-
2 \, |U_{\alpha1}|^2 \, |U_{\alpha2}|^2
\;.
\label{505}
\end{equation}
Combining Eqs.(\ref{504}) and (\ref{505}),
we obtain the following expression for
the survival probability,
\begin{equation}
P_{\nu_{\alpha}\to\nu_{\alpha}}
=
\left(
1
-
\sum_{k=3,4}
\left| U_{{\alpha}k} \right|^2
\right)^2
P_{\nu_{\alpha}\to\nu_{\alpha}}^{(1,2)}
+
\left(
\sum_{k=3,4}
\left| U_{{\alpha}k} \right|^2
\right)^2
\;,
\label{506}
\end{equation}
where
\begin{equation}
P_{\nu_{\alpha}\to\nu_{\alpha}}^{(1,2)}
=
1 -
{\displaystyle
2 \, |U_{\alpha2}|^2 \, |U_{\alpha1}|^2
\over\displaystyle
\left(
\sum_{j=1,2}
\left| U_{{\alpha}j} \right|^2
\right)^2
}
\left(
1
-
\cos
{\displaystyle
\Delta m^2_{21} \, L
\over\displaystyle
2 \, p
}
\right)
\label{507}
\end{equation}
is the survival probability
of $\nu_{\alpha}$
due to its mixing with
$\nu_1$ and $\nu_2$.
In Section \ref{SN3}
we have used Eq.(\ref{506}).

Equation (\ref{507})
can be written in terms
of a mixing angle
$\theta_{\alpha}$
defined by
\begin{equation}
\cos \theta_{\alpha}
=
{\displaystyle
|U_{\alpha1}|
\over\displaystyle
\sqrt{
\left(
\sum_{j=1,2}
\left| U_{{\alpha}j} \right|^2
\right)^2
}
}
\null \hskip2cm \null \displaystyle
\sin \theta_{\alpha}
=
{\displaystyle
|U_{\alpha2}|
\over\displaystyle
\sqrt{
\left(
\sum_{j=1,2}
\left| U_{{\alpha}j} \right|^2
\right)^2
}
}
\;.
\label{508}
\end{equation}
Then the expression (\ref{507})
for the survival probability
$ P_{\nu_{\alpha}\to\nu_{\alpha}}^{(1,2)} $
takes the standard two-generation form,
\begin{equation}
P_{\nu_{\alpha}\to\nu_{\alpha}}^{(1,2)}
=
1 - {1\over2}
\sin^2 2\theta_{\alpha}
\left(
1
-
\cos
{\displaystyle
\Delta m^2_{21} \, L
\over\displaystyle
2 \, p
}
\right)
\label{509}
\end{equation}

\begin{figure}[h]
\protect\caption{Exclusion regions in the
$ A_{\nu_{\mu};\nu_{e}} $--$ \Delta m^2 $
plane
for small
$ \left| U_{e1} \right|^2 $
and
$ \left| U_{\mu1} \right|^2 $
in the model with mixing of three neutrinos
discussed in Section \protect\ref{SN3}.
The regions excluded by
the BNL E776 and KARMEN
$ \nu_\mu \to \nu_e $
appearance experiments
are bounded by the dash-dotted and dash-dot-dotted curves,
respectively.
The dashed line represents
the results of the Bugey experiment.
The curve passing through the circles
is obtained
from the results
of the Bugey, CDHS and CCFR84 experiments
using Eq.(\protect\ref{226}).
The curve passing through the triangles
is obtained
from the results
of the Bugey, FNAL E531 and CCFR95 experiments
using Eq.(\protect\ref{228}).
The line passing through the squares
bounds the region that will be explored by
CHORUS and NOMAD.
The region allowed by the LSND experiment
is also shown as
the shadowed region limited by the two solid curves.}
\label{n3r1muel}
\end{figure}

\begin{figure}[h]
\protect\caption{Exclusion regions in the
$ A_{\nu_{e};\nu_{\tau}} $--$ \Delta m^2 $
for small
$ \left| U_{e1} \right|^2 $
and large
$ \left| U_{\mu1} \right|^2 $
in the model with mixing of three neutrinos
discussed in Section \protect\ref{SN3}.
The solid line represents
the results of the FNAL E531 experiment.
The dashed line represents
the results of the Bugey experiment.
The curve passing through the triangles
is obtained
from the results
of the Bugey, CDHS and CCFR84 experiments
using Eq.(\protect\ref{233}).
The curve passing through the squares
is obtained
from the results
of the CDHS, CCFR84 and BNL E776 experiments
using Eq.(\protect\ref{236}).
The curve passing through the circles
is obtained
from the results
of the BNL E776, FNAL E531 and CCFR95 experiments
using Eq.(\protect\ref{237}).}
\label{n3r2elta}
\end{figure}

\begin{figure}[h]
\protect\caption{Exclusion regions in the
$ A_{\nu_{\mu};\nu_{e}} $--$ \Delta m^2 $
plane
for small
$b_{e}$
and
$b_{\mu}$
in the model with mixing of four neutrinos
discussed in Section \protect\ref{SN4}.
The regions excluded by
the BNL E776 and KARMEN
$ \nu_\mu \to \nu_e $
appearance experiments
are bounded by the dash-dotted and dash-dot-dotted curves,
respectively.
The dashed line represents
the results of the Bugey experiment.
The curve passing through the circles
is obtained
from the results
of the Bugey, CDHS and CCFR84 experiments
using Eq.(\protect\ref{311}).}
\label{n4r1muel}
\end{figure}

\newpage

\begin{minipage}[h]{\textwidth}
\null\vskip-1cm
\begin{center}
\mbox{\epsfig{file=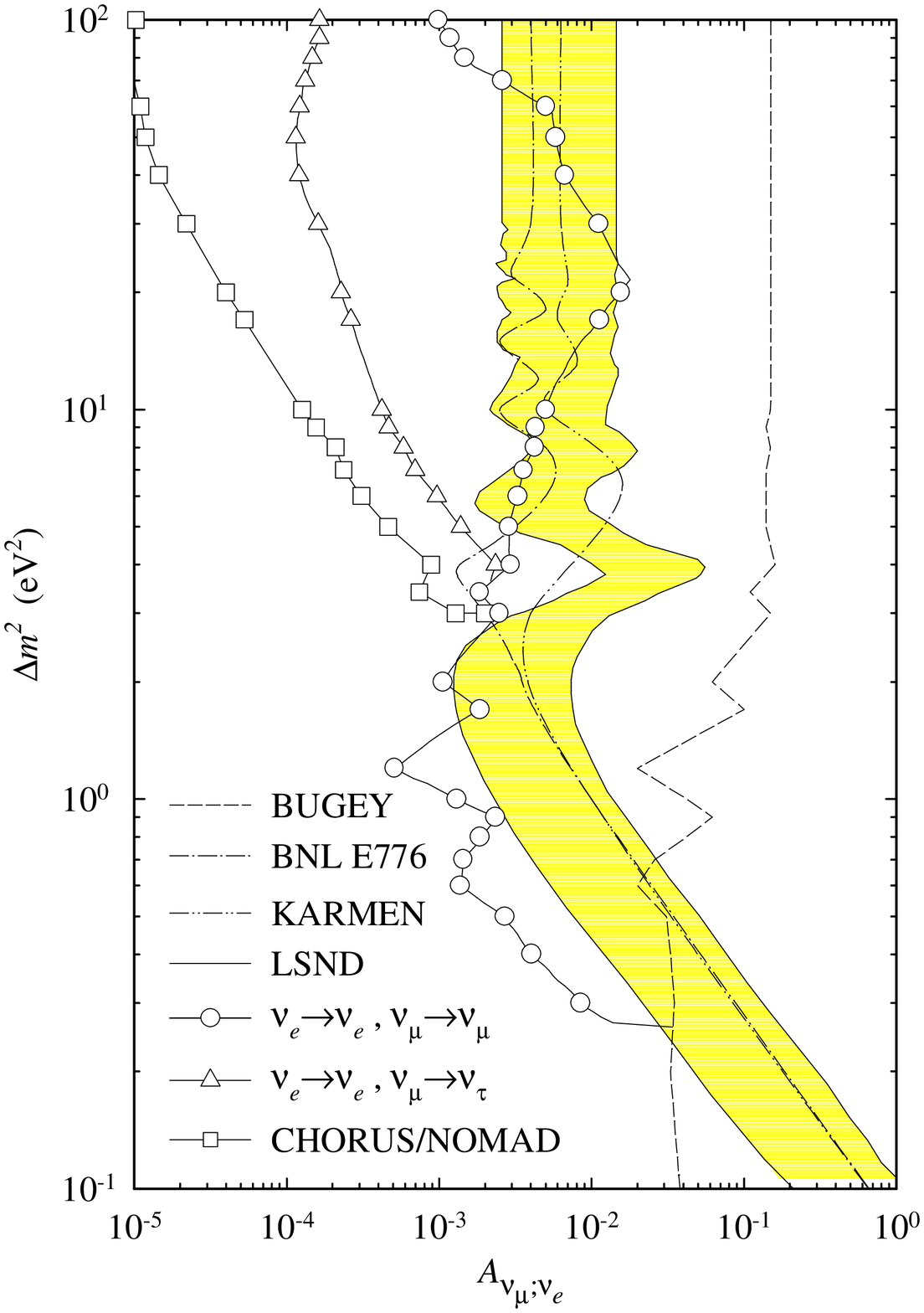,height=0.9\textheight}}
\end{center}
\end{minipage}
\vspace{1cm}
\begin{center}
{\Large Figure \ref{n3r1muel}}
\end{center}

\newpage

\begin{minipage}[h]{\textwidth}
\null\vskip-1cm
\begin{center}
\mbox{\epsfig{file=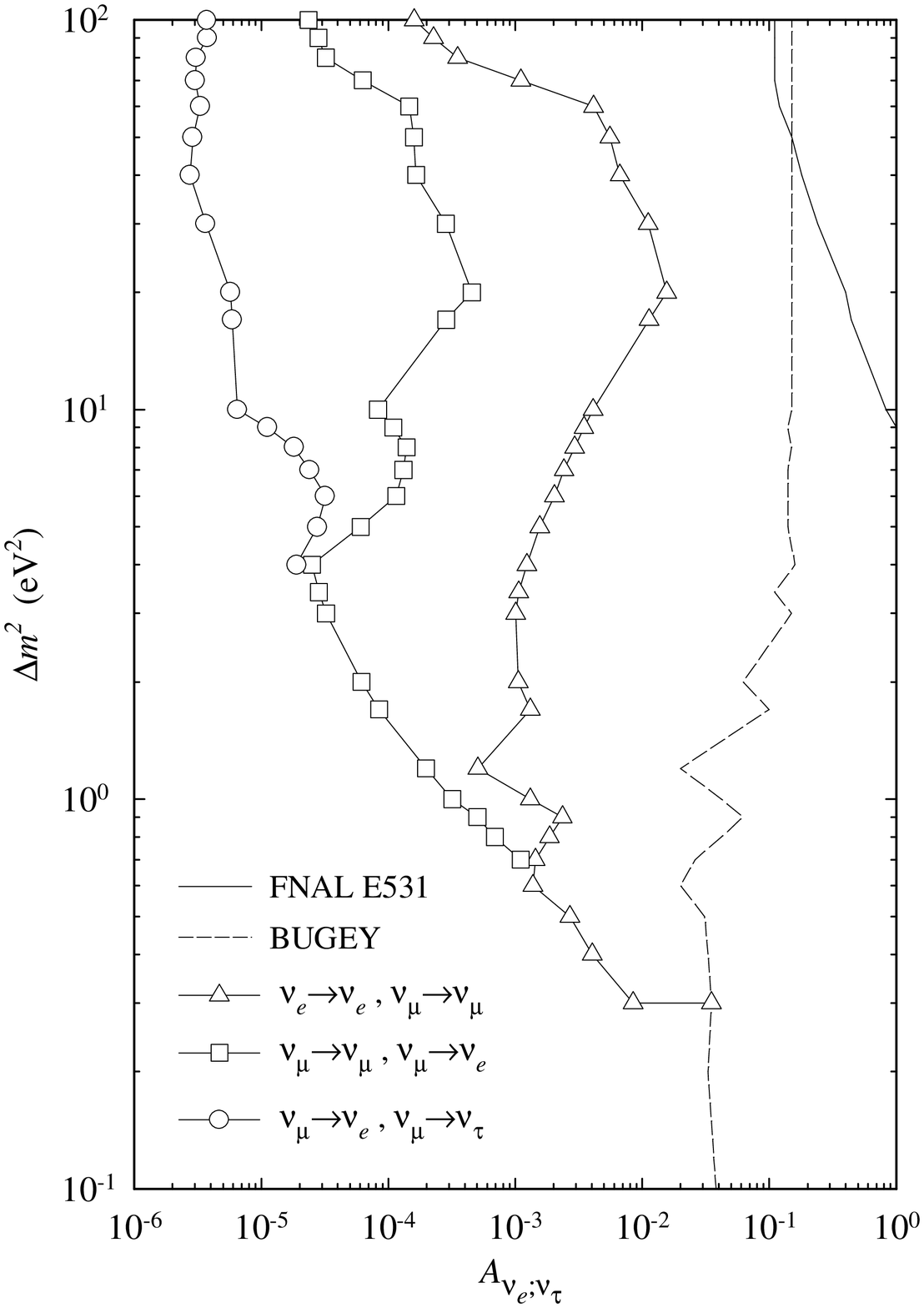,height=0.9\textheight}}
\end{center}
\end{minipage}
\vspace{1cm}
\begin{center}
{\Large Figure \ref{n3r2elta}}
\end{center}

\newpage

\begin{minipage}[h]{\textwidth}
\null\vskip-1cm
\begin{center}
\mbox{\epsfig{file=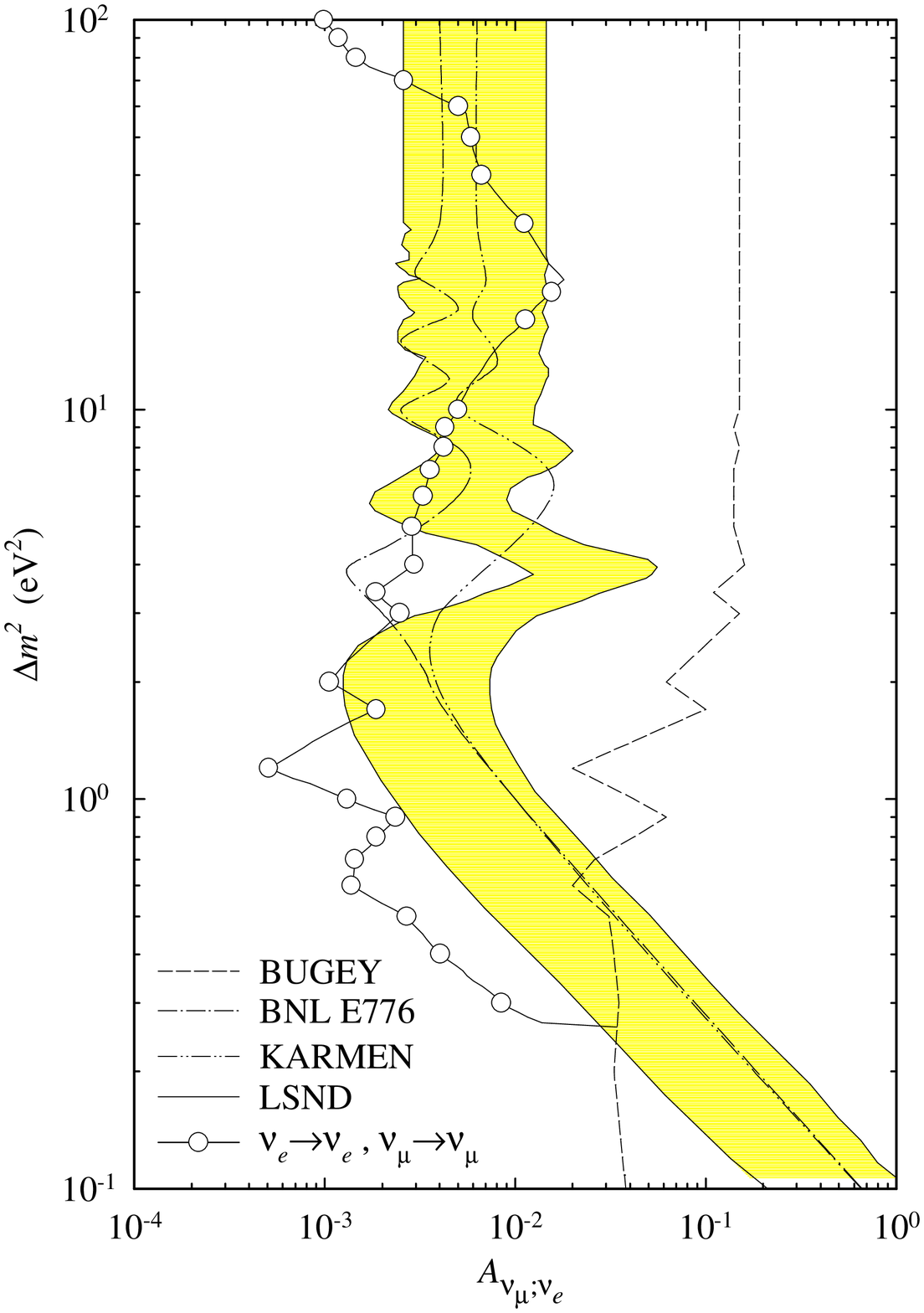,height=0.9\textheight}}
\end{center}
\end{minipage}
\vspace{1cm}
\begin{center}
{\Large Figure \ref{n4r1muel}}
\end{center}


\begin{references}

\bibitem{Homestake}
B.T. Cleveland et al.,
Nucl. Phys. B (Proc. Suppl.) {\bf 38}, 47 (1995).

\bibitem{Kamiokande}
K. S. Hirata et al.,
Phys. Rev. D {\bf 44}, 2241 (1991).

\bibitem{GALLEX}
GALLEX Coll.,
Phys. Lett. B {\bf 357}, 237 (1995).

\bibitem{SAGE}
V. Vermul,
Talk presented at the
{\it $7^{\mathrm{th}}$ International Workshop
on Neutrino Telescopes},
Venezia, February 1996.

\bibitem{bahcall}
J.N. Bahcall and R. Ulrich,
Rev. Mod. Phys. {\bf 60}, 297 (1988);
J.N. Bahcall,
{\it Neutrino Physics and Astrophysics},
Cambridge University Press, 1989;
J.N. Bahcall and M.H. Pinsonneault,
Rev. Mod. Phys. {\bf 64}, 885 (1992);
IASSNS-AST-95-24
(e-Print Archive: hep-ph/9505425).

\bibitem{turk}
S. Turck-Chi\`eze and I. Lopes,
Astrophys. J. {\bf 408}, 347 (1993);
S. Turck-Chi\`eze et al.,
Phys. Rep. {\bf 230}, 57 (1993).

\bibitem{cdf}
V. Castellani, S. Degl'Innocenti and G. Fiorentini,
Astronomy \& Astrophysics {\bf 271}, 601 (1993);
S. Degl'Innocenti,
Univ. of Ferrara preprint
INFN-FE-07-93.

\bibitem{GALLEXcal}
GALLEX Coll.,
Phys. Lett. B {\bf 342}, 440 (1995).

\bibitem{SAGEcal}
V.N. Gavrin,
Talk presented at the
{\it $7^{\mathrm{th}}$ International Workshop
on Neutrino Telescopes},
Venezia, February 1996.

\bibitem{phenomenological}
V. Castellani et al.,
Astron. Astrophys. {\bf 271}, 601 (1993);
S.A. Bludman et al.,
Phys. Rev. D {\bf 49}, 3622 (1994);
V. Berezinsky,
Comm. Nucl. Part. Phys. {\bf 21}, 249 (1994);
J.N. Bahcall,
Phys. Lett. B {\bf 338}, 276 (1994).

\bibitem{MSW}
S.P. Mikheyev and A.Yu. Smirnov,
Yad. Fiz. {\bf 42}, 1441 (1985)
[Sov. J. Nucl. Phys. {\bf 42}, 913 (1985)];
Il Nuovo Cimento C {\bf 9}, 17 (1986);
L. Wolfenstein,
Phys. Rev. D {\bf 17}, 2369 (1978);
Phys. Rev. D {\bf 20}, 2634 (1979).

\bibitem{SOLMSW}
GALLEX Coll.,
Phys. Lett. B {\bf 285}, 390 (1992);
X. Shi, D.N. Schramm and J.N. Bahcall,
Phys. Rev. Lett. {\bf 69}, 717 (1992);
P.I. Krastev and S.T. Petcov,
Phys. Lett. B {\bf 299}, 99 (1993);
N. Hata and P.G. Langacker,
Phys. Rev. {\bf 50}, 632 (1994);
L.M. Krauss, E. Gates and M. White,
Phys. Rev. D {\bf 51}, 2631 (1995);
G.L. Fogli and E. Lisi,
Astropart. Phys. {\bf 2}, 91 (1994);
G. Fiorentini et al.,
Phys. Rev. D {\bf 49}, 6298 (1994).

\bibitem{Pontecorvo}
B. Pontecorvo,
JETP (Sov. Fiz.) {\bf 53}, 1717 (1967)

\bibitem{BP78}
S.M. Bilenky and B. Pontecorvo,
Phys. Rep. {\bf 41}, 225 (1978).

\bibitem{SOLVAC}
P.I. Krastev and S.T. Petcov,
Phys. Lett. B {\bf 285}, 85 (1992);
Phys. Lett. B {\bf 299}, 99 (1993);
Phys. Rev. Lett. {\bf 72}, 1960 (1994);
V. Barger, R.J.N. Phillips, and K. Whisnant,
Phys. Rev. Lett. {\bf 69}, 3135 (1992).

\bibitem{KP95}
P.I. Krastev and S.T. Petcov,
Nucl. Phys. B {\bf 449}, 605 (1995).

\bibitem{KP96}
P.I. Krastev and S.T. Petcov,
Phys. Rev. D {\bf 53} (1996) 1665.

\bibitem{Kamiokande-atmospheric}
Y. Fukuda et al.,
Phys. Lett. B {\bf 335}, 237 (1994).

\bibitem{IMB}
R. Becker-Szendy et al.,
Nucl. Phys. B (Proc. Suppl.) {\bf 38}, 331 (1995).

\bibitem{Soudan}
M. Goodman,
Nucl. Phys. B (Proc. Suppl.) {\bf 38}, 337 (1995);
J. Schneps,
Talk presented at the
{\it $7^{\mathrm{th}}$ International Workshop
on Neutrino Telescopes},
Venezia, February 1996.

\bibitem{LSND}
C. Athanassopoulos et al.,
Phys. Rev. Lett. {\bf 75}, 2650 (1995).

\bibitem{hill}
J.E. Hill,
Phys. Rev. Lett. {\bf 75}, 2654 (1995).

\bibitem{KT89}
E.W. Kolb and M.S. Turner,
{\it The Early Universe},
Addison-Wesley, 1990.

\bibitem{CWKim}
C.W. Kim and A. Pevsner,
{\it Neutrinos in Physics and Astrophysics},
Contemporary Concepts in Physics, Vol. 8,
(Harwood Academic Press, Chur, Switzerland, 1993).

\bibitem{RPP}
Review of Particle Properties,
Phys. Rev. D {\bf 50}, 1173 (1994).

\bibitem{PS94}
S.T. Petcov and A.Yu. Smirnov,
Phys. Lett. B {\bf 322}, 109 (1994).

\bibitem{CM95}
D.O. Caldwell and R.N. Mohapatra,
Phys. Lett. B {\bf 354}, 371 (1995).

\bibitem{RS95}
G. Raffelt and J. Silk,
Phys. Lett. B {\bf 366}, 429 (1996).

\bibitem{CM93}
D.O. Caldwell and R.N. Mohapatra,
Phys. Rev. D {\bf 48}, 3259 (1993).

\bibitem{PV93}
J.T. Peltoniemi and J.W.F. Valle,
Nucl. Phys. B {\bf 406}, 409 (1993)

\bibitem{SSF93}
Z. Shi, D.N. Schramm and B.D. Fields,
Phys. Rev. D {\bf 48}, 2563 (1993).

\bibitem{PHKC95}
J.R. Primack et al.,
Phys. Rev. Lett. {\bf 74}, 2160 (1995).

\bibitem{GCGG95}
J.J. Gomez-Cadenas and M.C. Gonzalez-Garcia,
preprint CERN-TH/95-80
(e-Print Archive: hep-ph/9504246);
M.C. Gonzalez-Garcia,
preprint CERN-TH/95-285
(e-Print Archive: hep-ph/9510419).

\bibitem{SG95}
S. Goswami,
preprint CUPP-95/4 
(e-Print Archive: hep-ph/9507212).

\bibitem{r-process}
Y.Z. Qian et al.,
Phys. Rev. Lett. {\bf 71}, 1965 (1993);
Y.Z. Qian and G.M. Fuller,
Phys. Rev. D {\bf 51}, 1479 (1995);
G. Sigl, Phys. Rev. D {\bf 51}, 4035 (1995).

\bibitem{BP87}
S.M. Bilenky and S.T. Petcov,
Rev. Mod. Phys. {\bf 59}, 671 (1987).

\bibitem{Lisi}
G.L. Fogli, E. Lisi and D. Montanino,
Phys. Rev. D 49 (1994) 3626;
Astropart. Phys. {\bf 4}, 177 (1995);
G.L. Fogli, E. Lisi and G. Scioscia,
Phys. Rev. D {\bf 52}, 5334 (1995).

\bibitem{BBGK95}
S.M. Bilenky, A. Bottino, C. Giunti and C.W. Kim,
Phys. Lett. B {\bf 356}, 273 (1995).

\bibitem{BBGK96}
S.M. Bilenky, A. Bottino, C. Giunti and C.W. Kim,
preprint DFTT 2/96, JHU-TIPAC 96002
(e-Print Archive: hep-ph/9602216).

\bibitem{Minakata}
H. Minakata,
Phys. Lett. B {\bf 356}, 61 (1995);
Phys. Rev. D {\bf 52}, 6630 (1995);
O. Yasuda and H. Minakata,
preprint TMUP-HEL-9604
(e-Print Archive: hep-ph/9602386).

\bibitem{BPW95}
K.S. Babu, J.C. Pati and F. Wilczek,
Phys. Lett. B {\bf 359}, 351 (1995).

\bibitem{CF96}
C.Y. Cardall and G.M. Fuller,
preprint ASTROPH-9603071
(e-Print Archive: astro-ph/9602104). 

\bibitem{Bugey95}
B. Achkar et al.,
Nucl. Phys. B {\bf 434}, 503 (1995).

\bibitem{CDHS84}
F. Dydak et al.,
Phys. Lett. B {\bf 134}, 281 (1984).

\bibitem{CCFR84}
I.E. Stockdale et al.,
Phys. Rev. Lett. {\bf 52}, 1384 (1984).

\bibitem{SS92}
X. Shi and D.N. Schramm,
Phys. Lett. B {\bf 283}, 305 (1992).

\bibitem{troitsk}
A.I. Belesev et al.,
Phys. Lett. B {\bf 350}, 263 (1995).

\bibitem{mbb}
M. Doi et al.,
Phys. Lett. B {\bf 102}, 323 (1981);
L. Wolfenstein,
Phys. Lett. B {\bf 107}, 77 (1981).

\bibitem{MV94}
M. Moe and P. Vogel,
Annu. Rev. Nucl. Part. Sci. {\bf 44}, 247 (1994).

\bibitem{Moe95}
M.K. Moe,
Nucl. Phys. B (Proc. Suppl.) {\bf 38}, 36 (1995).

\bibitem{CPparity}
B. Kayser,
Phys. Rev. D {\bf 30}, 1023 (1984);
S.M. Bilenky, N.P. Nedelcheva and S.T. Petcov,
Nucl. Phys. B {\bf 247}, 61 (1984).

\bibitem{BNLE776}
L. Borodovsky et al.,
Phys. Rev. Lett. {\bf 68}, 274 (1992).

\bibitem{FNALE531}
N. Ushida
Phys. Rev. Lett. {\bf 57}, 2897 (1986).

\bibitem{CCFR95}
K.S. McFarland et al.,
Phys. Rev. Lett. {\bf 75}, 3993 (1995).

\bibitem{KARMEN}
B. Armbruster et al.,
Nucl. Phys. B (Proc. Suppl.) {\bf 38}, 235 (1995).

\bibitem{CHORUS}
G. Rosa,
Nucl. Phys. B (Proc.Suppl.) {\bf 40}, 85 (1995);
D. Macina,
Talk presented at TAUP 95
Toledo (Spain), Sept. 1995.

\bibitem{NOMAD}
A. Rubbia,
Nucl. Phys. B (Proc.Suppl.) {\bf 40}, 93 (1995);
M. Laveder,
Talk presented at TAUP 95
Toledo (Spain), Sept. 1995
(e-Print Archive: hep-ph/9601342).

\bibitem{COSMOS}
N.W. Reay et al.,
Fermilab proposal P803, Oct. 1993.

\bibitem{KP88}
P.I. Krastev and S.T. Petcov,
Phys. Lett. B {\bf 207}, 64 (1988). 

\bibitem{KP86}
T.K. Kuo and J. Pantaleone,
Phys. Rev. Lett. {\bf 57}, 1805 (1986);
Rev. Mod. Phys. {\bf 61}, 937 (1989).

\end{references}
\end{document}